\documentclass[11pt]{article}

\pdfoutput=1

\usepackage[T1]{fontenc}
\usepackage[latin9]{inputenc}
\usepackage[a4paper]{geometry}
\usepackage[active]{srcltx}
\usepackage{amsmath}
\usepackage{amssymb}
\usepackage{esint}
\usepackage{ulem}

\makeatletter

\usepackage{textcomp}

\pdfoutput=1 

\usepackage{jheppub}

\newcommand*\xbar[1]{%
  \hbox{%
    \vbox{%
      \hrule height 0.5pt 
      \kern0.3ex
      \hbox{%
        \kern-0.0em
        \ensuremath{#1}%
        \kern-0.0em
      }%
    }%
  }%
}

\usepackage{amsfonts}

\usepackage{stmaryrd}
\usepackage{mathtools}

\usepackage{natbib}
\usepackage{mathrsfs}
\usepackage{verbatim}

\usepackage{bbm}

\newcommand{\be}{\begin{equation}}
\newcommand{\ee}{\end{equation}}
\newcommand{\bea}{\begin{eqnarray}}
\newcommand{\eea}{\end{eqnarray}}

\renewcommand{\d}{\partial}

\title{\boldmath Logarithmic matching between past infinity and future infinity: The massless scalar field}

\author[a,b,c]{Oscar Fuentealba}
\author[c,d]{and Marc Henneaux}

\affiliation[a]{Instituto de Ciencias Exactas y Naturales (ICEN), Universidad Arturo Prat,\\ Playa Brava 3256, 1111346 Iquique, Chile}
\affiliation[b]{Facultad de Ciencias, Universidad Arturo Prat,\\ Avenida Arturo Prat Chac\'on 2120, 1110939 Iquique, Chile}
\affiliation[c]{Universit\'e Libre de Bruxelles and International Solvay Institutes, \\ ULB-Campus Plaine CP231, B-1050 Brussels, Belgium}
\affiliation[d]{Coll\`ege de France,  Universit\'e PSL, 11 place Marcelin Berthelot, \\ 75005 Paris, France}

\emailAdd{ofuentealba@unap.cl}
\emailAdd{marc.henneaux@ulb.be}

\preprint{}

\abstract{Matching conditions relating the fields at the future of past null infinity with the fields at the past of future null infinity play a central role in the analysis of asymptotic symmetries and conservation laws in asymptotically flat spacetimes. These matching conditions can be derived from initial data given on a Cauchy hypersurface by integrating forward and backward in time the field equations to leading order in an asymptotic expansion, all the way to future and past null infinities. The standard matching conditions considered in the literature are valid only in the case when the expansion near null infinity (which is generically polylogarithmic) has no dominant logarithmic term.  The absence of dominant logarithmic term, in turn,  holds only when the leading order of the initial conditions on a Cauchy hypersurface  (which contains no logarithm) fulfills definite parity conditions under the antipodal map of the sphere at infinity.  One can consistently consider opposite parity conditions.  While these do not conflict with the asymptotic symmetry group, they lead to a very different asymptotic behaviour near null infinity, where the expansion starts now with logarithmic terms that are no longer subdominant (even though such logarithmic terms are absent in the initial data), which implies different matching conditions.  

It turns out that many of the analytic features relevant to gravity are already present for massless spin zero and spin one fields. This paper is the first in a series in which we derive the matching conditions for a massless scalar field with initial conditions leading to logarithms at null infinity.  We prove that these involve the opposite sign with respect to the usual matching conditions. We also analyse the matching of the angle-dependent conserved charges that follow from the asymptotic decay and Lorentz invariance.  We show in particular that these are well defined and finite at null infinity even in the presence of leading logarithmic terms provided one uses the correct definitions.

The free massless scalar field has the virtue of presenting the polylogarithmic features in a particularly clear setting that shows their inevitability, since there is no subtle gauge fixing issue or nonlinear intrincacies involved in the problem. We also consider the case of higher spacetime dimensions where fractional powers of $r$ (odd spacetime dimensions) or subdominant logarithmic terms (even spacetime dimensions) are present.  Mixed matching conditions are then relevant. In subsequent papers, we will extend the analysis to the electromagnetic and the gravitational fields.}

\makeatother

\begin{document}
\maketitle \flushbottom

 \newpage{}

\section{Introduction}

The question of the behaviour of matter and gravitational fields in the vicinity of null infinity is a non-trivial dynamical question. From the perspective of the Cauchy problem, this question can be phrased as follows: given ``reasonable'' initial conditions on a Cauchy hypersurface, what type of expansions do the resulting fields admit near future null infinity, in particular, close to the ``critical sets" where spatial infinity and null infinity ``meet" \cite{Fried1,Friedrich:1999wk,Friedrich:1999ax,Friedrich:2002ru,Friedrich:2014rpa,Friedrich:2017cjg}\footnote{These critical sets correspond respectively to the past of future null infinity and the future of past null infinity and are very clearly described in \cite{Gasperin:2017apb,Paetz:2018nbd,Gasperin:2021vnm,Mohamed:2021rfg}.}? 

The reason that the question of the behaviour near null infinity is non-trivial is that spatial infinity and the critical sets define points with very different properties from the point of view of the relevant linear second order differential equations that govern the asymptotic dynamical evolution.  While spatial infinity corresponds to a regular point of these differential equations, the critical sets correspond by contrast to Fuchsian singularities. For $D=4$ and the leading order in the $1/r$ expansion, the roots of the indicial equation coincide.  One independent solution is then logarithm-free - actually, polynomial - and defines what is  called the ``$P$-branch".  The other independent solution involves logarithms and defines the `` $Q$-branch".  General solutions are linear combinations of both branches. As a result, even if the initial data obey standard expansions in $1/r$ at spatial infinity, the fields will generically develop polylogarithmic  terms of the form $r^{-k} \ln^m r$ near null infinity, as it is well documented (see \cite{Fried1,Friedrich:1999wk,Friedrich:1999ax,Friedrich:2002ru,Friedrich:2014rpa,Friedrich:2017cjg,Gasperin:2017apb,Paetz:2018nbd,Gasperin:2021vnm,Mohamed:2021rfg,Damour:1985cm,Christodoulou:1993uv,Chrusciel:1993hx,Chrusciel:1998he,Christodoulou:2000,ValienteKroon:2001pc,Valiente-Kroon:2002xys,ValienteKroon:2003ix,Bieri:2009xc,Henneaux:2018cst,Henneaux:2018gfi,Henneaux:2018hdj,Henneaux:2018mgn,Kehrberger:2021uvf,Gajic:2022pst,Minucci:2022hav,Bieri:2023cyn,Gasperin:2023uty,Kehrberger:2024clh,Sen:2024bax} and references therein).

These polylogarithmic terms  are not related to nonlinearities or to the presence of sources not vanishing at null infinity, or linked to a bad choice of gauge.  This is  particularly clear for a free massless scalar field $\phi$  in Minkowski space - a linear system with no gauge invariance.  One finds that if the scalar field behaves asymptotically at spatial infinity as $\phi \sim\frac{1 }{r}$, its behaviour at null infinity (in the vicinity of the critical sets)  is $\phi \sim\frac{\log r}{r}$ (rather than the expected $\phi \sim\frac{1 }{r}$), unless the leading $\frac{1 }{r}$-order of the initial data at spatial infinity is even under the antipodal map on the sphere.  In that case, the coefficient of the leading $\frac{\log r}{r}$-term at null infinity is zero, and the expansion starts with $\frac{1}{r}$, with subdominant terms of order $\frac{\log r}{r^2}$ and lower  \cite{Henneaux:2018mgn}. But if no parity condition is imposed on the leading  $\frac{1 }{r}$-order at spatial infinity, dominant terms of order  $\frac{\log r}{r}$ are unavoidable.  

The reason that parity conditions control the absence or presence of logarithms is due to the fact that the $P$-branch and the $Q$-branch obey opposite parity properties under the antipodal map of the hyperboloid at infinity, i.e., the combined antipodal map of the sphere at infinity with hyperbolic time reversal, which exchanges the past of future null infinity with the future of past null infinity.  The $P$-branch is even while the $Q$-branch is odd. Imposing even or odd parity conditions on the leading order of the initial data eliminates the dominant term of either the $Q$-branch or of the $P$-branch.

Similar features hold for electromagnetism or gravity where the leading terms in the expansion near null infinity  involve (in four spacetime dimensions) a dominant logarithmic term (e.g. of order $ \frac{\log r}{r^2}$ for the radial electric field rather than the expected order $\frac{1}{r^2}$) unless the initial conditions are appropriately constrained by (twisted) parity conditions \cite{Henneaux:2018cst,Henneaux:2018gfi,Henneaux:2018hdj}.  Even if one imposes the appropriate parity conditions that eliminate the leading logarithmic term, logarithms will appear in the expansion at subdominant orders.

The presence of polylogarithmic terms at null infinity is sometimes feared to conflict with the asymptotic symmetries.  We recall that the  asymptotic symmetry groups of electromagnetism and gravity are infinite-dimensional and equal either to the group of angle-dependent $U(1)$ transformations for electromagnetism \cite{He:2014cra,Kapec:2014zla,Kapec:2015ena} or the BMS group for gravity \cite{Bondi:1962px,Sachs:1962wk,Sachs:1962zza,Penrose:1962ij,Madler:2016xju,Alessio:2017lps,Ashtekar:2018lor}.  These groups, obtained first at future null infinity, were derived, however, for configurations that did not contain polylogarithmic terms. Such terms violate 
 the peeling properties of the Weyl tensor, thought to be crucial in the asymptotic context.  This raises the question as to whether these terms, when included as they should, spoil the BMS or $u(1)$ symmetries. 

The answer to this question is in fact obvious in the Hamiltonian description at spatial infinity and was given a few years ago in the work \cite{Henneaux:2018cst,Henneaux:2018gfi,Henneaux:2018hdj}:  the polylogarithmic terms that develop at null infinity do not conflict with the existence of the BMS symmetry or of the angle-dependent $u(1)$ symmetries, which can already be exhibited on the initial Cauchy hypersurfaces independently of the subsequent behaviour at null infinity.  This result has actually been confirmed directly at null infinity in  \cite{Geiller:2022vto,Geiller:2024ryw}. This strengthens the view already expressed by many authors that peeling is perhaps not so essential. 

Another question, is however still open and is the one on which this paper focuses. It is the question of matching past null infinity with future null infinity.

As we recalled, the asymptotic symmetry groups were derived first at future null infinity.  A similar analysis holds for past null infinity.  Since there is only one spacetime, the question arises then as to how to glue the transformations at future null infinity with those at past null infinity.  This is achieved through matching conditions that relate the leading order of the fields at the future of past null infinity with the leading order of the fields at the past of future null infinity, selecting a single ``diagonal" subgroup of asymptotic symmetries \cite{He:2014cra,Strominger:2013jfa,He:2014laa,Strominger:2017zoo}.  The matching conditions play a central role in the construction of the conserved charges following from the symmetry.

In deriving the matching conditions, an asymptotic form of the fields near null infinity that was free from logarithms was assumed in the works just cited.  One can easily show that if the terms involving the logarithms are subleading, the matching conditions, which relate the leading orders of the fields at past infinity and future infinity,  are unchanged.  This was shown in \cite{Henneaux:2018gfi,Henneaux:2018hdj}. 

But what happens for initial conditions leading to dominant logarithmic terms at null infinity? While some parity conditions are necessary to get a well-defined action principle and a well-defined Hamiltonian of the asymptotic symmetry groups, these need not be of the type that eliminate the leading logarithmic terms at null infinity.  Alternative parity conditions have been considered and are perfectly consistent from the Hamiltonian viewpoint \cite{Henneaux:2018cst,Henneaux:2018gfi,Henneaux:2018hdj}.   They yield dominant logarithmic terms at null infinity, necessitating a re-analysis of the matching conditions.

The purpose of this paper is to generalize the matching conditions to cover this dominant logarithmic case, leading to what we call ``logarithmic matching between past infinity and future infinity".  We consider first the case of a free massless scalar field, for which the question of the matching conditions for the fields raises the same types of problems as for the spin-$1$ and spin-$2$ cases, but in a simpler linear context, free from gauge-fixing subtleties.

One key ingredient in the analysis is to determine the behaviour of the Klein-Gordon solutions near null infinity  for general initial conditions on a Cauchy hypersurface (and in particular in the vicinity of the critial sets where null infinity and spatial infinity "touch").  The general solution for arbitrary initial data was given  in  \cite{Henneaux:2018mgn}.  We go beyond this  earlier work by (i) deriving the matching conditions in all cases, including the one when leading logarithmic terms are present at null infinity (ii) studying the behaviour of the conserved charges and showing that they are well behaved in all cases, including again the case when leading logarithmic terms are present at null infinity.   The analysis of the charges in that latter case is more subtle because the expression for the charge at null infinity is then different from the expression valid in the absence of such leading logarithmic terms, which, if used, would lead to incorrect conclusions.  The correct expressions can be derived by carefully following the Hamiltonian expressions from the initial Cauchy hypersurface all the way to null infinity.

We also extend the analysis to higher dimensions, where the general solutions to the equations of motion can be found in \cite{Henneaux:2018mgn} (appendix A, see also \cite{Henneaux:2019yqq}).  Again the relevant second order linear differential equation has Fuchsian singularities at the critical sets but the indicial roots now do not coincide. The general solution involves again a singular $Q$-branch that develops logarithms at null infinity in even spacetime dimensions (the indicial roots differ by an integer) and fractional powers of $r$ in odd spacetime dimensions (the indicial roots do not differ by an integer).  Contrary to the situation prevailing for $D=4$, both branches can now be simultaneously kept in the Hamiltonian description, leading to mixed matching conditions.  We discuss also the behaviour of the charges and find again no problem at null infinity for generic initial conditions if one adopts the correct definitions of the charges. This analysis is particularly instructive not only because it indicates why the charges, which are obviously well-defined at spatial infinity, remain well-defined as we go to the critical sets in spite of the non-analytical terms that emerge there, but also because it shows how to extract the charges from the intricate expressions of the fields at null infinity.

In subsequent papers, we shall cover the matching conditions and the behaviour of the charges at null infinity for Maxwell and Einstein theories.  We shall pay special attention to  the matching of the gauge potential and the metric components (and not just the field strength or Weyl tensor), which involve the important Goldstone bosons associated with the breaking of the angle-dependent asymptotic symmetries. Since the gauge potential and the metric are not invariant under gauge transformations, this is a problem that must be handled with care and actually needs convenient choices of the gauge fixing conditions to get results where the matching is transparent.

Our paper is organized as follows.  We start in Section {\bf \ref{Sec:4DH}} with a brief survey of the behaviour of a massless scalar field in 4 spacetime dimensions near spatial infinity, and exhibit two independent sets of angle-dependent asymptotic quantities (conserved charges).   We then proceed in Section {\bf \ref{Sec:4D}} with a detailed analysis of the behaviour of the general solution of the free massless Klein-Gordon equation in 4 spacetime dimensions as one goes toward null infinity.  We recall in particular how logarithms appear in that limit from logarithm-free initial data and relate the leading behaviour  near null infinity to parity properties of the leading terms in the asymptotic expansion of the initial data near spatial infinity.  We also provide an in-depth discussion of the conserved charges and provide their expression at null infinity, which is finite in all cases.  We then turn, in Section {\bf \ref{Sub:Matching}}, to the matching conditions and derive the matching conditions relevant for solutions with a leading logarithmic term, which involves a crucial minus sign with respect to the standard matching conditions.  Section {\bf \ref{SecHighD}} is devoted to the generalization to higher spacetime dimensions.  A key new feature is that parity conditions on the leading orders at spatial infinity are not necessary any more, so that both the leading $P$-branch and the leading $Q$-branch must be considered simultaneously, leading to mixed matching conditions.  To emphasize the differences between even and odd spacetime dimensions, we consider explicitly the cases of $D=5$ and $D=6$.  We conclude our paper in Section {\bf \ref{sec:Conclusions}}, with general comments and prospects for future work. Finally, two appendices treat more technical points: useful coordinate systems (Appendix {\bf \ref{app-decomp}}) and some explicit formulas on the Gegenbauer/ultraspherical functions of the second kind (Appendix {\bf \ref{App:Gegenbauer}}).

To close this introduction, we stress that the logarithms present in the $Q$-branch are not to be confused with the logarithmic terms induced by logarithmic gauge transformations or supertranslations \cite{Fuentealba:2022xsz,Fuentealba:2023syb,Fuentealba:2023rvf,Fuentealba:2023huv}, which are already present in the initial data through specific improper gauge transformations.  Our initial conditions for the scalar field are free from logarithms.

 \section{Behaviour of a scalar field at spatial infinity in $4$ spacetime dimensions}
\label{Sec:4DH}

We discuss in this section the free massless  Klein-Gordon field  in four spacetime dimensions,
\be
\Box \phi = -\partial_t^2 \phi + \triangle \phi = 0\,.
\ee
As explained in the introduction, this case is particularly simple and strikingly demonstrates the unavoidability of logarithmic terms at null infinity.  Indeed, the equation of motion is linear and there is no gauge invariance and hence no difficulties with gauge conditions.  Hence, one cannot blame the appearance of logarithms on nonlinearities or gauge fixing problems.  The scalar field in four dimensions was explicitly treated in \cite{Henneaux:2018mgn}.  We go beyond the analysis in that reference by generalizing the matching conditions to the case when unusual parity conditions are fulfilled by the leading orders.

\subsection{Conserved charges}

We will assume that on any constant-$t$ slice, the massless scalar field $\phi$ decays as
\be
\phi = \frac{\xbar \phi}{r} + \frac{\phi^{(2)}}{r^2}+ \mathcal O(r^{-3})\, , \label{Eq:decay0a}%
\ee
where the coefficients of the powers of $r^{-1}$ in this expansion are functions of the angles $x^A$.
The long range nature of $\phi$ mimics that of gauge potentials and is crucial for the appearance of logarithms at null infinity.  For $\phi$ of compact support, there would be no logarithm \cite{Wald:1984rg} (Proposition 11.1.1).

It has been observed in \cite{Henneaux:2018mgn} that 
the leading $1/r$ term in $\phi$ define conserved quantities, 
\be
\partial_t \xbar \phi (x^A)= 0 \, .
\ee
 This merely follows from the assumed asymptotic behaviour and Lorentz invariance.  The property  holds even in the presence of Lorentz-invariant interactions compatible with the above decay. 

This conservation law is implemented in the Hamiltonian formulation by assuming that the conjugate momentum $\pi = \dot \phi$ behaves at infinity (in Cartesian coordinates)\footnote{We recall that $\pi$ is a spatial density and picks up a factor of $r^2 \sqrt{\gamma}$ in polar coordinates.  In this paper, we will denote by $\xbar \pi$ the leading order ($\mathcal O(r^{-2})$) in the expansion of $\pi$ in Cartesian coordinates.} as
\be
\pi = \frac{\xbar \pi}{r^2} + \frac{\pi^{(3)}}{r^3}+ \mathcal O(r^{-4})\,, \label{Eq:decay0b}
\ee
which is indeed compatible with Lorentz invariance under which $\phi$ transforms as $\delta_b \phi =( b_i x^i) \pi $ (for boosts).

In the free case, there is another infinite set of asymptotic charges that follows from the Klein-Gordon equation
$
\partial_t^2 \phi = \triangle \phi
$,
namely
\be
 \partial_t^2 \phi^{(2)} = 0\,,
\ee 
since $\triangle \phi =  \mathcal O(r^{-3})$ (we always assume when needed that our $\mathcal O(r^k)$ quantities are such that $\partial_i \mathcal O(r^k) = \mathcal O(r^{k-1})$). This is equivalent to
\be
\partial_t \xbar \pi (x^A) = 0 \, .
\ee

We thus have an infinite number of conserved charges, namely $\xbar \phi (x^A)$ and $\xbar \pi (x^A)$.
Having no accompanying  bulk term,  these charges do not have well-defined functional derivatives and hence are not  well-defined canonical generators.  
The situation is improved in the dual formulation of the scalar field in terms of a $2$-form, where the  infinite number of angle-dependent charges can be interpreted as generators, \`a la Regge-Teitelboim \cite{Regge:1974zd},  of the improper gauge symmetries \cite{Benguria:1976in} of the dual $2$-form \cite{Henneaux:2018mgn,Campiglia:2018see,Francia:2018jtb,Nguyen:2022nnx}.   We shall not elaborate on this point here.

It is useful to combine the conserved charges  to form the surface integral
\be
Q[\epsilon, \eta] = \oint (\xbar \epsilon\, \xbar \pi +\xbar \zeta\, \xbar \phi ) \sqrt{\xbar \gamma} d^2 x\,,  \label{Eq:Charge00}
\ee
over the $2$-sphere at infinity on the $t=$ constant spacelike hyperplanes.  Here, $\sqrt{\xbar \gamma}$ is  the determinant of the unit round metric on the $2$-sphere (see Appendix {\bf \ref{app-decomp}} for conventions), while $\xbar \epsilon(x^A)$ and $\xbar \zeta(x^A)$ are arbitrary functions on the $2$-sphere, which would be the parameters of the transformation generated by $Q$ if $Q$ was a well-defined generator, but which are here merely useful bookkeeping devices.  The surface integral $Q[\epsilon, \eta]$ is conserved provided $\dot {\xbar \epsilon} = 0 = \dot {\xbar \zeta}$.  One can expand the parameters and charges in terms of spherical harmonics if one so wishes.

\subsection{Parity conditions}

It turns out that the decay (\ref{Eq:decay0a})-(\ref{Eq:decay0b}) does not guarantee that the action 
\be
S[\phi, \pi] = \int dt \int d^3 x \left[ \pi \dot{\phi} - \frac12 \left(\pi^2 + (\partial_k \phi)^2\right) \right]
\ee
is well-defined.  Indeed, the kinetic term $\int d^3x \pi \dot{\phi}$ generically diverges logarithmically.  To get a finite action,  one needs to impose extra conditions on the leading orders $\xbar \phi$ and $\xbar \pi$  \cite{Henneaux:2018mgn}. 

\subsubsection*{Standard boundary conditions}
 One possibility is to impose that $\xbar \phi$ is even under the antipodal map on the sphere, symbolically written as $x^A \rightarrow -x^A$, while  $\xbar \pi$ is odd.  This makes the kinetic term finite since the coefficient of the logarithmic term vanishes. It is the standard choice developed in \cite{Henneaux:2018mgn}.  

\subsubsection*{Unusual boundary conditions }
An unusual but equally consistent choice is to impose that  $\xbar \phi$ is odd while  $\xbar \pi$ is even. This alternative possibility also makes the logarithmic divergence in the kinetic term absent.

Both choices, which lead to very distinct behaviours near null infinity, are compatible with Poincar\'e invariance, and both choices will be pursued in our analysis.

\section{Behaviour of a scalar field at null infinity in $4$ spacetime dimensions}
\label{Sec:4D}

\subsection{Going to null infinity}
 
\subsubsection{Klein-Gordon equation in hyperbolic coordinates}
In order to understand how the development of the initial data (\ref{Eq:decay0a})-(\ref{Eq:decay0b}) behaves near null infinity, one needs to solve the massless Klein-Gordon equation,
\be
\Box \phi = 0\,.
\ee
We will perform that task without imposing any parity condition first, and will  impose these conditions on the general solution at the very end.

Following \cite{Ashtekar:1978zz,BeigSchmidt,Beig:1983sw,Compere:2011ve}, we go to hyperbolic coordinates, where the wave equation separates order by order.  These coordinates are reviewed in Appendix {\bf \ref{app-decomp}}.

Since the slice $s=0$ (with $s$ the hyperbolic time) coincides with the hyperplane $t = 0$, on which $\eta = r$, we expand the scalar field in powers of $\eta$ as,
\be
\phi = \sum_{k \geq 0} \frac{\phi^{(k)}(x^a)}{\eta^{k + 1}}\,,
\ee
with  $ \xbar \phi (x^A) = \phi^{(0)}(s=0, x^A)$ etc. This is useful because each order obeys independent equations.  

The wave equation for $\phi$ becomes indeed in hyperbolic coordinates
\begin{equation}
    \d_\mu(\sqrt{-g} g^{\mu\nu} \d_\nu \phi) = \eta \sqrt {-h}
    \Big(\eta^{-1} \d_\eta(\eta^3
    \d_\eta \phi) + \mathcal D^a \mathcal D_a \phi \Big)= 0\,,
\end{equation}
where $\mathcal D_a$ is the covariant derivative with respect to the metric $h_{ab}$ and $\mathcal D^a = h^{ab} \mathcal D_b$.  It thus decouples order by order in $\eta$ as
\be
\mathcal D^a \mathcal D_a \phi^{(k)}  + (k + 1)(k-1) \phi^{(k)} = 0\,.
\ee
Expanding the hyperboloid wave operator $\mathcal D^a \mathcal D_a$ yields then explicitly
\be
(1-s^2) \partial^2_s \phi^{(k)}  - \xbar D_A \xbar D^A \phi^{(k)} - \frac{(k^2 -1)}{1-s^2} \phi^{(k)} = 0\,.
\ee

We now use spherical symmetry and expand the scalar fields $\phi^{(k)}$ in spherical harmonics, including the factor $(1-s^2)^{\frac{1-k}{2}}$ that eliminates the double pole in the coefficient of the undifferentiated term in the equation,
\be
\phi^{(k)}(s, x^A) = (1-s^2)^{\frac{1-k}{2}} \sum_{l\geq 0} \sum_{\vert m \vert \leq l}  \Theta^{(k)}_{lm}(s) Y_{lm}(x^A) \, ,  \quad \xbar D_A \xbar D^A Y_{lm}= - l(l+1) Y_{lm} \, ,
\ee
to obtain
\be
(1-s^2) \partial^2_s \Theta_{lm}^{(k)}  + 2 ( k - 1) s \partial_s  \Theta_{lm}^{(k)} + \left[l(l+1)-k(k-1)\right] \Theta_{lm}^{(k)} = 0 \, .   \label{Eq:FuchsD=4}
\ee
Note that we do NOT restrict the value of $l$ to be smaller than or equal to $k$.  The integer $l$ is unrestricted.  Our approach is thus more general than the one of \cite{Gasperin:2023uty}, which imposes $l \leq k$.  From now on, the notation $\sum_{l,m}$ will denote the $l$-unrestricted sum $ \sum_{l\geq 0} \sum_{\vert m \vert \leq l}$.

The equation (\ref{Eq:FuchsD=4}) is a second order linear differential equation with Fuchsian singularities at $\pm 1$ (and $\infty$).  The indicial equation for the exponent $\alpha$ appearing in the solutions  is
\be
\alpha (\alpha-k) = 0
\ee
at $\pm 1$, with roots $\alpha =0$ and $\alpha = k$ that differ by an integer (see for instance Section 7.2.2 of \cite{Petrini:2019ujm} for a general discussion of linear second order differential equations).

\subsubsection*{Case $k=0$}

As shown in \cite{Henneaux:2018mgn}, the first term in the expansion ($k=0$) brings the dominant logarithmic term.  We consider it separately for that reason.

For $k = 0$, the equation reduces to the standard Legendre equation.  Therefore the coefficients $\Theta_{lm}^{(0)}(s)$ are linear combinations of the Legendre polynomials $P_l^{(\frac12)}(s)$ and the Legendre functions of the second kind $Q_l^{(\frac12)}(s)$,
\be
\Theta_{lm}^{(0)}(s) = \Theta_{lm}^{P(0)}P_l^{(\frac12)}(s) +  \Theta_{lm}^{Q(0)}Q_l^{(\frac12)}(s)\,,
\ee
where $ \Theta_{lm}^{P(0)}$ and $ \Theta_{lm}^{Q(0)}$ are arbitrary constants.   Thus,
\be
\phi^{(0)}(s, x^A) = \phi^{(0)}_P(s, x^A) + \phi^{(0)}_Q(s, x^A) \,,
\ee
where the $P$-branch and $Q$-branch are respectively given by
\be
\phi^{(0)}_P(s, x^A) = (1-s^2)^{\frac{1}{2}} \sum_{l,m} \Theta^{P(0)}_{lm}P_l^{(\frac12)}(s) Y_{lm}(x^A) \, , 
\ee
and
\be
\phi^{(0)}_Q(s, x^A) = (1-s^2)^{\frac{1}{2}} \sum_{l,m} \Theta^{Q(0)}_{lm}Q_l^{(\frac12)}(s) Y_{lm}(x^A) \, .
\ee
The superscript ${}^{(\frac12)}$ has been added to $P_l^{(\frac12)}(s)$ and $Q_l^{(\frac12)}(s)$ for uniformity reasons that will be clear in the next subsection.

We recall that the Legendre polynomials are finite and take the value $1$ at $s=1$,
\be
P_l^{(\frac12)}(s) = 1 + \mathcal O(1-s)\,,
\ee
while the Legendre functions $Q_l^{(\frac12)}(s)$ have a logarithmic singularity,
\be
Q_0^{(\frac12)}(s) = \frac12 \log\left(\frac{1+s}{1-s}\right) \, \quad Q_l^{(\frac12)}(s) = P_l^{(\frac12)}(s) Q_0^{(\frac12)}(s) + R_l^{(\frac12)}(s)\,, \label{Eq:LF2}
\ee
where $R_l^{(\frac12)}(s)$ is a definite polynomial of degree $l-1$, so that
\be
Q_l^{(\frac12)}(s) = -\frac12 \log(1-s) +  \mathcal O(1) \,,
\ee
near $s=1$.

We furthermore note that $P_l^{(\frac12)}(-s)= (-1)^l P_l^{(\frac12)}(s)$ and $Q_l^{(\frac12)}(-s)=-(-1)^l Q_l^{(\frac12)}(s)$ so that the $P$-branch is even under the combined $s$-reversal ($s \rightarrow -s$) and antipodal map of the $2$-sphere, symbolically denoted $x^A \rightarrow - x^A$, while the $Q$-branch is odd,
\be
\phi^{(0)}_P(-s, -x^A) =\phi^{(0)}_P(s, x^A) \, , \qquad \phi^{(0)}_Q(-s, -x^A) =-\phi^{(0)}_Q(s, x^A) \, .
\ee

\subsubsection*{Case $k>0$}
The solutions for $k>0$ are also  explicitly given in  \cite{Henneaux:2018mgn} and recalled in this subsection.

When $l\geq k$ they can be expressed in terms of Gegenbauer's (or ultraspherical) polynomials $P^{(\lambda)}_n(s)$ and Gegenbauer (or ultraspherical) functions of the second kind $Q^{(\lambda)}_n(s)$, where we have set $\lambda = k + \frac12$ and $n= l-k$ to match the standard conventions.  The behaviour of $P^{(\lambda)}_n(s)$ and $Q^{(\lambda)}_n(s)$ near the pole $s=1$ of the differential equation is, to leading order,
\be P^{(\lambda)}_n(s) \sim \begin{pmatrix}n+2\lambda -1 \\ n \end{pmatrix}  \, , \qquad Q^{(\lambda)}_n(s) \sim \frac{(1-s^2)^{- \lambda + \frac12}}{2\lambda-1} = \frac{(1-s^2)^{-k}}{2\lambda-1}  \, ,
\ee
($n \geq 0$, $ \lambda > \frac12$).  While the next terms in the polynomials $P^{(\lambda)}_n(s)$ are smooth and go to zero as $s \rightarrow 1$, the next terms in $Q^{(\lambda)}_n(s)$ involve subdominant (with respect to $(1-s^2)^{-k}$) logarithmic terms.  For instance, 
$$ Q^{(\frac32)}_0(s) = \frac12 \, \frac{s}{1-s^2} + \frac14 \log \left(\frac{1+s}{1-s}\right) \, .$$
The ultraspherical  solutions obey also definite parity properties,
 \be
P_n^{(\lambda)}(-s)= (-1)^n P_n^{(\lambda)}(s) \, , \qquad Q_n^{(\lambda)}(-s)=-(-1)^n Q_n^{(\lambda)}(s) \, . \label{Eq:ParityUltra}
\ee

In terms of the Gegenbauer polynomials and functions of the second kind,  the $k>0$ solutions can be expressed as:
\be
\phi^{(k)}(s, x^A) = \phi^{(k)}_P(s, x^A) + \phi^{(k)}_Q(s, x^A) \, ,
\ee
where the $P$-branch and $Q$-branch are respectively given by
\be
\phi^{(k)}_P(s, x^A) = (1-s^2)^{\frac{1-k}{2}} \sum_{l,m} \Theta^{P(k)}_{lm}\widetilde P_{l-k}^{(k+\frac12)}(s) Y_{lm}(x^A) \, , 
\ee
and
\be
\phi^{(k)}_Q(s, x^A) = (1-s^2)^{\frac{1-k}{2}} \sum_{l,m} \Theta^{Q(k)}_{lm}\widetilde Q_{l-k}^{(k+ \frac12)}(s) Y_{lm}(x^A) \, ,
\ee
with
\be
\widetilde P_{n}^{(k+\frac12)}(s) = (1-s^2)^{k} P^{(k+\frac12)}_n(s)\, , \qquad \widetilde Q_{n}^{(k+\frac12)}(s) = (1-s^2)^{k} Q^{(k+\frac12)}_n(s)\, .
\ee
The $P$-branch is associated with the root $\alpha = k$ while the $Q$-branch is associated with the root $\alpha = 0$. One has 
\be
\lim_{s \rightarrow 1} \widetilde P_{n}^{(k+\frac12)}(s) = 0 \, , \qquad \lim_{s \rightarrow 1} \widetilde Q_{n}^{(k+\frac12)}(s) =  \frac{1}{2k} \,,
\ee
and
\be
\phi^{(k)}_P(-s, -x^A) =(-1)^k \phi^{(k)}_P(s, x^A) \, , \quad \phi^{(k)}_Q(-s, -x^A) =- (-1)^k \phi^{(k)}_Q(s, x^A) \, . \label{Eq:ParityX7}
\ee 
We stress again, to emphasize their inevitability, that the $Q$-branch involves for $k>0$ subdominant logarithmic terms, such as $(1-s) \log (1-s)$.

When $l< k$ ($n<0$), one gets similar expressions where both $\widetilde P_{n}^{(k+\frac12)}(s)$ and $\widetilde Q_{n}^{(k+\frac12)}(s)$ are now polynomials, which have been chosen to obey the parity conditions (\ref{Eq:ParityUltra}).  The differential equation has indeed only polynomials solutions. Being polynomials, these solutions have some finite  values as $s \rightarrow 1$, which are non-zero \cite{Henneaux:2018mgn}.  

Instead of the set $\{\widetilde P_{n}^{(k+\frac12)}(s), \widetilde Q_{n}^{(k+\frac12)}(s)\}$ which has definite parity properties, one could alternatively take   $\{\overline P_{n}^{(k+\frac12)}(s), \widetilde Q_{n}^{(k+\frac12)}(s)\}$ as set of independent solutions (when $n<0$), where 
$\overline P_{n}^{(k+\frac12)}(s)$ is the linear combination of $\widetilde P_{n}^{(k+\frac12)}(s)$ and $\widetilde Q_{n}^{(k+\frac12)}(s)$ that vanishes at $s=1$.  It does so as $(1-s)^k$, corresponding to the indicial root $\alpha = k$.  The polynomials $\overline P_{n}^{(k+\frac12)}(s)$ do not have definite parity properties.  The $\overline P$-terms in the asymptotic expansion will be subdominant at null infinity with respect to the $Q$-terms.

\subsubsection{Friedrich coordinates - Bondi coordinates}
\label{SubSec:FB}

While useful for integrating the equations of motion for initial data given in a $\frac{1}{\eta}$ expansion, the hyperbolic coordinates badly cover null infinity.  Indeed, as one tends to null infinity along a radial null geodesics, e.g., for $\mathscr{I^+}$, 
\be
t = r + b \qquad (b <0 )
\ee
one finds that for $r \rightarrow \infty$, $s$ and $\eta$ always behave as $s \rightarrow 1$ and $\eta \rightarrow \infty$, no matter what the null geodesic is.  The limiting values of $(s, \eta)$ are always $(1, \infty)$ so that the information about the null geodesic (i.e., $b$) and where it reaches $\mathscr{I^+}$ is lost.  This suggests to go to new coordinates that do not suffer from this limitation.  

These coordinates were introduced by Friedrich  \cite{Fried1,Friedrich:1999wk,Friedrich:1999ax,Friedrich:2002ru,Friedrich:2014rpa} and are well reviewed in \cite{Gasperin:2017apb,Gasperin:2021vnm,Mohamed:2021rfg}.  They are described in detail in  Appendix {\bf \ref{app-decomp}}.

In Friedrich coordinates, the scalar field reads
\be 
\phi = \phi_P + \phi_Q\,,
\ee
with
\be
\phi_P = (1-s^2) \sum_{k\geq0, l \geq 0, \vert m\vert \leq l} \rho^{-1-k}\Theta^{P(k)}_{lm}\widetilde P_{l-k}^{(k+\frac12)}(s) Y_{lm}(x^A) \, , 
\ee
and
\be
\phi_Q = (1-s^2)  \sum_{k\geq0, l \geq 0, \vert m\vert \leq l} \rho^{-1-k} \Theta^{Q(k)}_{lm}\widetilde Q_{l-k}^{(k+ \frac12)}(s) Y_{lm}(x^A) \, .
\ee
Since null infinity corresponds to taking the limit $s \rightarrow \pm 1$ while keeping $\rho$ fixed, and since all the $\widetilde P_{l-k}^{(k+\frac12)}(s) $ and $\widetilde Q_{l-k}^{(k+\frac12)}(s)$ are bounded in that limit except $\widetilde Q_{l}^{\frac12)}(s)$ which diverges only logarithmically,   we see that the scalar field goes to zero as we go to null infinity, as expected (for $k=0$, $(1-s) \log (1-s) \rightarrow 0$).

We then proceed to Bondi coordinates $(u,r)$, related to $(s,\rho)$ as follows,
\be
s = 1 + \frac{u}{r}\, , \quad \rho = - 2u - \frac{u^2}{r} \, , \quad 1-s^2 = -2 \, \frac{u}{r} -\frac{u^2}{r^2}\,.
\ee
Here, $u<0$ since we are interested in the limit  of going to the past of future null infinity.

Writing the solution $\phi$ in Bondi coordinates, we get
\begin{eqnarray}
&&\phi_P =\frac{1}{r} \sum_{l,m} \Theta^{P(0)}_{lm} Y_{lm}(x^A)  \nonumber \\
&& \qquad +  \frac{1}{r}  \sum_{k>0}  \sum_{l<k}  \sum_{\vert m\vert \leq l} \frac{1}{(-2u)^k}\Theta^{P(k)}_{lm}\widetilde P_{l-k}^{(k+\frac12)}(1) Y_{lm}(x^A) + \mathcal O\left(\frac{1}{r^{2}} \right) 
\,,\, \qquad 
\end{eqnarray}
since $P_{l}^{(\frac12)}(1)=1$ and $\widetilde P_{l-k}^{(k+\frac12)}(1) =0$ for $l\geq k$ ($k>0$), as well as
\begin{eqnarray}
\phi_Q  &=&\left(\frac{1}{r}+ \mathcal O(r^{-2})\right) \sum_{l,m} \Theta^{Q(0)}_{lm} \left( P_{l}^{(\frac12)}(1)Q_0^{(\frac12)}(1+\frac{u}{r}) +R_{l}^{(\frac12)}(1) \right) Y_{lm}(x^A)  \nonumber \\
&& +  \frac{1}{r}  \sum_{k>0,l,m} \frac{1}{(-2u)^k}\Theta^{Q(k)}_{lm}\widetilde Q_{l-k}^{(k+\frac12)}(1) Y_{lm}(x^A) + \mathcal O\left(\frac{\log r}{r^{2}} \right)\,,
\end{eqnarray}
where we have expressed $Q_l^{(\frac12)}$ in terms of $Q_0^{(\frac12)}$ through  (\ref{Eq:LF2}) in order to display more transparently the asymptotic behaviour.

Let us analyse first the $P$-branch. All values of $k\geq 0$ contribute at order $1/r$ at null infinity, even though they correspond to terms of order $1/r^{k+1}$ at spatial infinity.  However, as we go to the past of future null infinity, these $1/r$ contributions goes to zero like $(-2u)^{-k}$.  Therefore, only  the $k=0$ contribution survives at  $\mathscr{I^+_-}$.  This $k=0$ contribution is $$\frac{1}{r} \sum_{l,m} \Theta^{P(0)}_{lm} Y_{lm}(x^A)$$  and in fact it does not depend on $u$.   It defines an infinite number of angle-dependent conserved quantities, which will be connected to the conserved quantities defined at spatial infinity in Subsection  {\bf \ref{Sub:Matching}}.  We can rewrite the $P$-branch at null infinity as
\be
\phi_P = \frac{\Phi(u, x^A)}{r} + \mathcal O \left(\frac{1}{r^2} \right)\,,  
\ee
with
\be
\lim_{u \rightarrow - \infty}  \Phi(u, x^A) =  \Phi(x^A)=  \sum_{l,m} \Theta^{P(0)}_{lm} Y_{lm}(x^A)\,,  \label{Eq:SchemeP}
\ee
involving only the $k=0$ contribution.

To analyse the $Q$-branch, we expand $Q_0^{(\frac12)}(s) = Q_0^{(\frac12)}(1+u/r) = (1/2)\big(\log(r/u) + \log(2+u/r)\big)$ (see  (\ref{Eq:LF2})) as
\be
Q_0^{(\frac12)}(1+u/r) = \frac12\Big(\log r  - \log(-u) + \log 2 \Big) + \mathcal O (r^{-1})\,,
\ee
and use explicitly $P_l^{(\frac12)}(1) = 1$ to get
\begin{eqnarray}
\phi_Q &=&\frac12 \frac{\log r}{r} \sum_{l,m} \Theta^{Q(0)}_{lm}  Y_{lm}(x^A)  \nonumber \\
 && + \frac{1}{r}  \sum_{l,m} \Theta^{Q(0)}_{lm} \left(  \frac12\Big(  - \log(-u) + \log 2 \Big) + R_{l}^{(\frac12)}(1) \right)  Y_{lm}(x^A)  \nonumber \\
&& +  \frac{1}{r}  \sum_{k>0,l,m} \frac{1}{(-2u)^k}\Theta^{Q(k)}_{lm}\widetilde Q_{l-k}^{(k+\frac12)}(1) Y_{lm}(x^A) + \mathcal O\left(\frac{\log r}{r^{2}} \right)\,.
\end{eqnarray}
There is now a leading $\frac{\log r}{r}$ contribution, originating entirely from the $k=0$ term.  It does not depend on $u$ and defines also an infinite number of angle-dependent conserved quantities. There are then $\frac{1}{r}$ contributions originating from all values of $k$, but only the $k=0$ term contributes in the limit $u \rightarrow - \infty$ and involves a $\log(-u)$ divergence.  This divergence in the coefficient of the $1/r$ term goes hand in hand with the coefficient of the $\frac{\log r}{r}$-term (they are present or absent simultaneously).   Finally there are subdominant polylogarithmic terms originating from all values of $k$.

The presence of coefficients that diverge as $u \rightarrow - \infty$ might seem disturbing but it should be recalled that the Bondi coordinates cover badly the past of future null infinity, which is missed, in fact, by taking the limit $u \rightarrow - \infty$ while keeping $r$ fixed. This limit is equivalent to keeping $r$ fixed and taking $t \rightarrow - \infty$, and amounts thus to go  to past infinity, which is outside the scope of this paper.  To study the limit at the critical sets, which is the object of this work, one must use the Friedrich coordinates to conclude that $\phi$ unambiguously goes to zero there, even if it does involve non analytical terms.

To parallel the expression (\ref{Eq:SchemeP}) for the $P$-branch,  we can rewrite the $Q$-branch at null infinity as
\be
\phi_Q = \frac12 \frac{\log r}{r} \Psi (u, x^A) + \mathcal O \left(\frac{1}{r} \right)\,,
\ee
where
\be
\Psi (u, x^A) =  \Psi(x^A) =  \sum_{l,m} \Theta^{Q(0)}_{lm} Y_{lm}(x^A)\,,
\label{Eq:SchemeQ}
\ee
is $u$-independent and involves only the $k=0$ contribution.

\subsection{Connection with Hamiltonian conserved quantities}

We can easily connect the conserved quantities exhibited at null infinity to the conserved quantities found at spatial infinity because we have at hand the form of the general solution. One simply sets $s=0$ in the general solution expressed in hyperbolic coordinates  since $s=0$ is equivalent to $t=0$.  

\subsubsection{$P$-branch}
This yields for the $P$-branch, noting that $\pi \equiv \dot \phi = \frac{1}{r} \partial_s \phi$ on $t=0$,
\begin{eqnarray}
\phi_P(t=0) &=& \frac{1}{r} \xbar \phi(x^A)   \nonumber \\  && +  \sum_{k  \in 2 \mathbb{N}+1} \frac{1}{r^{k+1}} \left(\sum_{l \in 2 \mathbb{N}+1, m}\Theta^{P(k)}_{lm}\widetilde P_{l-k}^{(k+\frac12)}(0) Y_{lm}(x^A)\right) \nonumber \\
&& + \sum_{k \in 2 \mathbb{N}+2} \frac{1}{r^{k+1}} \left(\sum_{l  \in 2 \mathbb{N}, m}\Theta^{P(k)}_{lm}\widetilde P_{l-k}^{(k+\frac12)}(0) Y_{lm}(x^A) \right)\,,
\end{eqnarray}
with
\be
\xbar \phi (x^A) = \sum_{l  \in 2 \mathbb{N}, m}\Theta^{P(0)}_{lm} P_{l}^{(\frac12)}(0) Y_{lm}(x^A)  \,,
\ee
and, for $\xbar \pi$ (in Cartesian coordinates)
\begin{eqnarray}
\pi_P(t=0) &=& \frac{1}{r^2} \xbar \pi(x^A)   \nonumber \\  && +  \sum_{k  \in 2 \mathbb{N}+1} \frac{1}{r^{k+2}} \left(\sum_{l \in 2 \mathbb{N}, m}\Theta^{P(k)}_{lm}\partial_s\widetilde P_{l-k}^{(k+\frac12)}(0) Y_{lm}(x^A)\right) \nonumber \\
&& + \sum_{k \in 2 \mathbb{N}+2} \frac{1}{r^{k+2}} \left(\sum_{l  \in 2 \mathbb{N}+1, m}\Theta^{P(k)}_{lm}\partial_s \widetilde P_{l-k}^{(k+\frac12)}(0) Y_{lm}(x^A) \right)\,, \qquad
\end{eqnarray}
with
\be
\xbar \pi(x^A) = \sum_{l  \in 2 \mathbb{N}+1, m}\Theta^{P(0)}_{lm}\partial_s  P_{l}^{(\frac12)}(0) Y_{lm}(x^A)\,,
\ee
where we have explicitly used the fact that due to the parity properties of the ultraspherical polynomials, $\widetilde P_n(s=0)$ vanishes for $n$ odd, while $\partial_s \widetilde P_n(s=0)$ vanishes for $n$ even.

We thus see that on Cauchy hypersurfaces, the coefficient  $\xbar \phi$ of the leading ($1/r$) order  of the scalar field is even under the sphere antipodal map for the $P$-branch, while the coefficient $\xbar \pi$ of the leading ($1/r^2$) order of its conjugate momentum is odd.   These angle-dependent conserved quantities  are completely determined by -- and completely determine in turn -- the coefficients $\Theta^{P(0)}_{lm}$.  Their knowledge is therefore equivalent to (\ref{Eq:SchemeP}). 

\subsubsection{$Q$-branch}
For the $Q$-branch, we get similar expressions,
\begin{eqnarray}
\phi_Q(t=0) &=& \frac{1}{r}\xbar \phi (x^A)  \nonumber \\  && +  \sum_{k  \in 2 \mathbb{N}+1} \frac{1}{r^{k+1}} \left(\sum_{l \in 2 \mathbb{N}, m}\Theta^{Q(k)}_{lm}\widetilde Q_{l-k}^{(k+\frac12)}(0) Y_{lm}(x^A)\right) \nonumber \\
&& + \sum_{k \in 2 \mathbb{N}+2} \frac{1}{r^{k+1}} \left(\sum_{l  \in 2 \mathbb{N}+1, m}\Theta^{Q(k)}_{lm}\widetilde Q_{l-k}^{(k+\frac12)}(0) Y_{lm}(x^A) \right)\,,\qquad
\end{eqnarray}
with
\be
\xbar \phi (x^A) = \sum_{l  \in 2 \mathbb{N}+1, m}\Theta^{Q(0)}_{lm} Q_{l}^{(\frac12)}(0) Y_{lm}(x^A)\,,
\ee
and
\begin{eqnarray}
\pi_Q(t=0) &=& \frac{1}{r^2} \xbar \pi(x^A)   \nonumber \\  && +  \sum_{k  \in 2 \mathbb{N}+1} \frac{1}{r^{k+2}} \left(\sum_{l \in 2 \mathbb{N}+1, m}\Theta^{Q(k)}_{lm}\partial_s\widetilde Q_{l-k}^{(k+\frac12)}(0) Y_{lm}(x^A)\right) \nonumber \\
&& + \sum_{k \in 2 \mathbb{N}+2} \frac{1}{r^{k+2}} \left(\sum_{l  \in 2 \mathbb{N}, m}\Theta^{Q(k)}_{lm} \partial_s \widetilde Q_{l-k}^{(k+\frac12)}(0) Y_{lm}(x^A) \right)\,,\qquad
\end{eqnarray}
with 
\be
\xbar \pi(x^A) =\sum_{l  \in 2 \mathbb{N}, m}\Theta^{Q(0)}_{lm}\partial_s Q_{l}^{(\frac12)}(0) Y_{lm}(x^A)\,,
\ee
where this time, $\widetilde Q_n(s=0)$ vanishes for $n$ even, while $\partial_s \widetilde Q_n(s=0)$ vanishes for $n$ odd.
Note that there is {\bf no} logarithm in the expansion of the $Q$-branch on Cauchy hypersurfaces but that  the coefficient  $\xbar \phi$ of the leading ($1/r$) order  of the scalar field is now odd under the antipodal map for the $Q$-branch, while the coefficient $\xbar \pi$ of the leading ($1/r^2$) order of its conjugate momentum is now even.  The apparently benign change in the parity conditions of the initial data that does not affect the $1/r$ dependence has dramatic consequences in the behaviour at null infinity as it controls the appearance of the $\log r$ terms..

As for the $P$-branch, these angle-dependent conserved quantities $\xbar \phi$ and $\xbar \pi$ are completely determined by -- and completely determine -- the coefficients $\Theta^{Q(0)}_{lm}$.  Their knowledge is therefore equivalent to (\ref{Eq:SchemeQ}).

\subsubsection{Conserved charges and parity conditions}

We now impose the parity conditions on the leading orders of the field and its conjugate momentum, necessary to have a well-defined canonical formulation.

\subsubsection*{Standard boundary conditions}
In the standard case, one imposes that $\xbar \phi$ is even under the antipodal map $x^A \rightarrow - x^A$ on the sphere, while  $\xbar \pi$ is odd.  This forces $\Theta^{Q(0)}_{lm}= 0$ and selects the $P$ branch for the leading order.   There is therefore no leading logarithmic term at infinity and $\Psi(x^A) = 0$.

The charge $\Phi(x^A)$ of Eq (\ref{Eq:SchemeP}) can be expressed in terms of the initial data on the Cauchy hypersurface $t = 0$ as 
\be
 \Phi(x^A)=  \sum_{l \in 2 \mathbb{N},m} \frac{ \xbar{\phi}_{lm}}{P_{l}^{(\frac12)}(0)} Y_{lm}(x^A) + \sum_{l \in 2 \mathbb{N} +1,m} \frac{\xbar{\pi}_{lm}}{\partial_s  P_{l}^{(\frac12)}(0)} Y_{lm}(x^A)\,,
\ee
where $\xbar{\phi}_{lm}$ and $\xbar{\pi}_{lm}$ are respectively the coefficients of $ Y_{lm}(x^A)$ in the spherical harmonic expansion of $\xbar \phi$ and $\xbar \pi$.
We thus see that both the even function $\xbar \phi$ and the odd function $\xbar \pi$ combine to form a single conserved function $\Phi(x^A)$ of the angles, with no definite parity under the antipodal map at future null infinity.

The general solution compatible with the standard parity conditions considered in this subsection reads, at null infinity,
\begin{eqnarray}
\phi &=&\frac{1}{r} \Phi(x^A)  +  \frac{1}{r}  \sum_{k>0,l<k,m} \frac{1}{(-2u)^k}\Theta^{P(k)}_{lm}\widetilde P_{l-k}^{(k+\frac12)}(1) Y_{lm}(x^A)  \nonumber \\
&&+  \frac{1}{r}  \sum_{k>0,l,m} \frac{1}{(-2u)^k}\Theta^{Q(k)}_{lm}\widetilde Q_{l-k}^{(k+\frac12)}(1) Y_{lm}(x^A) + \mathcal O\left(\frac{\log r}{r^{2}} \right)   \label{Eq:PBranchLeading}
\end{eqnarray}
from which we infer three important points:
\begin{itemize}
\item While the  quantities $\xbar \phi$ and $\xbar \pi$ are conserved as we progress on spacelike Cauchy hypersurfaces, the function $\Phi(u, x^A)$ to which they connect at null infinity is not conserved along null infinity since $\Phi$ depends on $u$ through the subleading contributions from the $P$  and $Q$ branches in (\ref{Eq:PBranchLeading}), which need not be set to zero.  These represent the radiation that ``escape" through null infinity.  It is given by a mixture of $P$ and $Q$ modes. 
\item While there is no leading logarithmic term in the development  (\ref{Eq:PBranchLeading}), there are subleading ones coming from the ($k>0$, $n\geq 0$)-terms  in the expansion of the $Q$-branch.  These terms are unavoidable unless we impose, on top of the condition $\Theta^{Q(0)}_{lm}= 0$, the somewhat artificial requirement  $\Theta^{Q(k)}_{lm}= 0$ for $k>0$, $n \geq 0$.   
\item The expansion (\ref{Eq:PBranchLeading}) coincides to leading order with the ones generally considered in the literature (see for instance \cite{Satishchandran:2019pyc}) but differs from them already at order $\mathcal O\left(\frac{\log r}{r^{2}} \right)$.  From the point of view of general initial conditions on Cauchy hypersurfaces, these other expansions are therefore somewhat limited.
\end{itemize}

\subsubsection*{Unusual boundary conditions }

We now impose the opposite parity conditions that $\xbar \phi$ is odd under the antipodal map $x^A \rightarrow - x^A$ on the sphere, while  $\xbar \pi$ is even.  This forces $\Theta^{P(0)}_{lm}= 0$ and selects the $Q$ branch for the leading order.   There is therefore a leading logarithmic term at infinity and $\Psi(x^A) \not= 0$.  While this term is generally avoided by most authors at null infinity, it leads to no particular complication at spatial infinity and, as stressed in  \cite{Henneaux:2018mgn}, a perfectly consistent Hamiltonian description can be provided with these opposite parity conditions.

We can write $\Psi(x^A)$ of Eq (\ref{Eq:SchemeQ}) in terms of the initial data on the Cauchy hypersurface $t = 0$ as 
\be
 \Psi(x^A)= \sum_{l \in 2 \mathbb{N} +1,m} \frac{\xbar{\phi}_{lm}}{  Q_{l}^{(\frac12)}(0)} Y_{lm}(x^A) + \sum_{l \in 2 \mathbb{N},m} \frac{ \xbar{\pi}_{lm}}{\partial_s Q_{l}^{(\frac12)}(0)} Y_{lm}(x^A) 
\ee
where $\xbar{\phi}_{lm}$ and $\xbar{\pi}_{lm}$ are again the coefficients of $ Y_{lm}(x^A)$ in the spherical harmonic expansion of $\xbar \phi$ and $\xbar \pi$, respectively.
Both the odd function $\xbar \phi$ and the even function $\xbar \pi$ combine to form a single conserved function $\Psi(x^A)$ of the angles, with no definite parity under the antipodal map.  

The general solution compatible with the non-standard parity conditions considered in this subsection reads, at null infinity,
\begin{eqnarray}
\phi &=&\frac12 \frac{\log r}{r} \Psi(x^A)  \nonumber \\
 && + \frac{1}{r}  \sum_{l,m} \Theta^{Q(0)}_{lm} \left(  \frac12\Big(  - \log(-u) + \log 2 \Big) + R_{l}^{(\frac12)}(1) \right)  Y_{lm}(x^A)  \nonumber \\
&& +   \frac{1}{r}  \sum_{k>0,l<k,m} \frac{1}{(-2u)^k}\Theta^{P(k)}_{lm}\widetilde P_{l-k}^{(k+\frac12)}(1) Y_{lm}(x^A) \nonumber \\
&& +  \frac{1}{r}  \sum_{k>0,l,m} \frac{1}{(-2u)^k}\Theta^{Q(k)}_{lm}\widetilde Q_{l-k}^{(k+\frac12)}(1) Y_{lm}(x^A) + \mathcal O\left(\frac{\log r}{r^{2}} \right) \, .
\end{eqnarray}
The coefficient $\Psi(x^A)$ of the leading term is this time absolutely conserved all along null infinity since there is no other $\frac{\log r}{r}$ contribution, in a way very similar to the Newman-Penrose charges in general relativity \cite{Newman:1965ik,Kroon:1998tu}.

\subsection{More on the conserved charges}

It is instructive to track more precisely the behavior of the conserved quantities as we go to null infinity.

The conserved quantities (\ref{Eq:Charge00}) are defined on the Cauchy hypersurfaces of constant $t$.  How are we going to define them in the hyperbolic coordinates?  Since $s= 0$ coincide with the Cauchy hypersurface $t=0$, and since
\be
\partial_s \phi \Bigr\rvert_{t=0} = \frac{\partial t}{\partial s} \Bigr\rvert_{t= 0} \frac{\partial \phi}{\partial t}\Bigr\rvert_{t=0}
\Rightarrow \xbar \pi = \partial_s \phi^{(0)}\ee
(with $\phi^{(0)}$ the coefficient of the $\eta^{-1}$-term in the expansion in hyperbolic coordinates), one can rewrite $Q$ at $s=0$ as
\be
Q[\xbar \epsilon, \xbar \zeta ]\Bigr\rvert_{s=0} =  \oint (\xbar \epsilon \partial_s{ \phi^{(0)}} + \xbar \zeta \,\phi^{(0)} ) \sqrt{\gamma} d^2 x  \Biggr\rvert_{s= 0}\,.
\ee
If we then introduce a function $\epsilon^{(0)}(s, x^A)$ such that $\epsilon^{(0)}\Bigr\rvert_{t=0} = \xbar \epsilon$ and $\partial_s\epsilon^{(0)}\Bigr\rvert_{t=0} = - \xbar \zeta$, the conserved quantities become
\be
Q[ \epsilon^{(0)}]\Bigr\rvert_{s=0} =  \oint ( \epsilon^{(0)} \partial_s{ \phi^{(0)}} - \partial_s{ \epsilon^{(0)}} \,\phi^{(0)} ) \sqrt{\gamma} d^2 x  \Biggr\rvert_{s= 0} \, .\label{Eq:QinS}
\ee

This rewriting of the conserved quantity suggests how one should define $Q$ for other values of $s$ in the hyperbolic slicing.  Since the slices $\vert s \vert < 1$ are spacelike allowing no flux at infinity, we naturally request $\frac{dQ}{ds} = 0$, which amounts to adopt the same expression (\ref{Eq:QinS}) for all $s$ 
\be
Q[ \epsilon^{(0)}] =  \oint ( \epsilon^{(0)} \partial_s{ \phi^{(0)}} - \partial_s{ \epsilon^{(0)}} \,\phi^{(0)} ) \sqrt{\gamma} d^2 x \,, \label{Eq:QinHyp}
\ee
and require that $ \epsilon^{(0)}$ obeys the same equation as $ \phi^{(0)}$
\be
(1-s^2) \partial^2_s \epsilon^{(0)}  - \xbar D_A \xbar D^A \epsilon^{(0)} +\frac{1}{1-s^2} \epsilon^{(0)} = 0 \, .
\ee

The requirement that $\frac{dQ}{ds} = 0$ is very natural. If one could view the conserved charges as generators of a symmetry, one could justify their form from the knowledge of the transformations that they must generate.  This independent justification is not available here, but is available in the electromagnetic and gravitational cases, where the charges (completed by bulk terms proportional to the constraints) have a well-defined action as generators of the asymptotic symmetry.

By the same derivations as the ones followed for $ \phi^{(0)}$, we then get
\be
(1-s^2) \partial^2_s \Xi_{lm}^{(0)}  - 2 s \partial_s  \Xi_{lm}^{(0)} + l(l+1)) \Xi_{lm}^{(0)} = 0 \, ,
\ee
with
\be
\epsilon^{(0)}(s, x^A) = (1-s^2)^{\frac{1}{2}} \sum_{l,m} \Xi^{(0)}_{lm}(s) Y_{lm}(x^A) \, ,
\ee
leading to 
\be
\epsilon^{(0)}(s, x^A) = \epsilon^{(0)}_P(s, x^A) + \epsilon^{(0)}_Q(s, x^A) \,,
\ee
with
\be
\epsilon^{(0)}_P(s, x^A) = (1-s^2)^{\frac{1}{2}} \sum_{l,m} \Xi^{P(0)}_{lm}P_l^{(\frac12)}(s) Y_{lm}(x^A) \, , 
\ee
and
\be
\epsilon^{(0)}_Q(s, x^A) = (1-s^2)^{\frac{1}{2}} \sum_{l,m} \Xi^{Q(0)}_{lm}Q_l^{(\frac12)}(s) Y_{lm}(x^A) \, .
\ee

If we plug these expressions in the formula (\ref{Eq:QinHyp}) giving the charge, we find four terms, 
\be
Q = Q_{P,P} + Q_{P,Q} + Q_{Q,P} + Q_{Q,Q}
\ee
with
\begin{eqnarray}
Q_{P,P} &=& \oint\left(\epsilon^{(0)}_P\partial_{s}\phi^{(0)}_P-\partial_{s}\epsilon^{(0)}_P\phi^{(0)}_P\right)\sqrt{\xbar\gamma}d^{2}x \, , \\
Q_{P,Q} &=& \oint\left(\epsilon^{(0)}_P\partial_{s}\phi^{(0)}_Q-\partial_{s}\epsilon^{(0)}_P\phi^{(0)}_Q\right)\sqrt{\xbar\gamma}d^{2}x \, ,\\
Q_{Q,P} &=& \oint\left(\epsilon^{(0)}_Q\partial_{s}\phi^{(0)}_P-\partial_{s}\epsilon^{(0)}_Q\phi^{(0)}_P\right)\sqrt{\xbar\gamma}d^{2}x  \, ,\\
Q_{Q,Q} &=& \oint\left(\epsilon^{(0)}_Q\partial_{s}\phi^{(0)}_Q-\partial_{s}\epsilon^{(0)}_Q\phi^{(0)}_Q\right)\sqrt{\xbar\gamma}d^{2}x  \, .
\end{eqnarray}
It is easy to check by direct computation that both $Q_{P,P}$ and $Q_{Q,Q}$ vanish
in the limit $s \rightarrow 1$ due to cancellations between the two terms in the integral, which have the same structure.  Only $Q_{P,Q}$ and $Q_{Q,P}$ can be different from zero.

\subsubsection*{Standard boundary conditions}

The integral $Q_{Q,P}$ gives the charge relevant for the standard boundary conditions characterized by $\phi_{Q}^{(0)}=0$ (pure $P$-branch). One might fear that it diverges at the critical set $s\rightarrow1$ because of the logarithm present in the Legendre function of the second kind:
\begin{equation}
\lim_{s->1}Q_{l}^{(\frac{1}{2})}(s)=-\frac{1}{2}\log(1-s)+q_{l}^{(0)}+\mathcal{O}(1-s)\,.
\end{equation}
However one can verify that the logarithms cancel each other between  $\epsilon^{(0)}_Q\partial_{s}\phi^{(0)}_P$ and $\partial_{s}\epsilon^{(0)}_Q\phi^{(0)}_P$. The remaining finite
part that comes from $\epsilon_{Q}^{(0)}\partial_{s}\phi^{(0)}_P$ is
zero at $s=1$, while the finite part from $-\partial_{s}\epsilon_{Q}^{(0)}\phi^{(0)}_P$
does not vanish at $s=1$ because
$\lim_{s\rightarrow1}(1-s^{2})\partial_{s}Q=1$.
Hence, we derived the relation:
\begin{equation}
\lim_{s\rightarrow1}Q_{Q,P}=-\oint\left(\sum_{l,m}\Xi_{lm}^{Q(0)}Y_{lm}(x^{A})\right)\left(\sum_{l',m'}\Theta_{l'm'}^{P(0)}Y_{l'm'}(x^{A})\right)\sqrt{\xbar\gamma}d^{2}x\,.
\end{equation}

\subsubsection*{Non-standard boundary conditions}

For the non-standard boundary conditions $\phi_{P}^{(0)}=0$ (pure $Q$-branch), it is $Q_{P,Q}$ that gives the charge.

The expression for $Q_{P,Q}$ is identical to (minus) that for $Q_{Q,P}$ with the roles of $\epsilon^{(0)}$ and $\phi^{()}$ exchanged, so that we can immediately infer
\begin{equation}
\lim_{s\rightarrow1}Q=\oint\left(\sum_{l,m}\Xi_{lm}^{P(0)}Y_{lm}(x^{A})\right)\left(\sum_{l',m'}\Theta_{l'm'}^{Q(0)}Y_{l'm'}(x^{A})\right)\sqrt{\xbar\gamma}d^{2}x\,.
\end{equation}

\vspace{.2cm}

The charges are well defined in the limit $s \rightarrow 1$ for both the $P$-branch and the $Q$-branch, as it should since they are conserved.  The mechanism by which finiteness is maintained through compensations between $\epsilon$ and $\phi$ is similar to the one discussed in \cite{Henneaux:2019yqq}.  The expression for the charges also exactly picks up the coefficient $\Phi(x^A)$ of (\ref{Eq:SchemeP}) of the $1/r$ term at null infinity for the $P$-branch and the coefficient $\Psi(x^A)$ of (\ref{Eq:SchemeQ}) of the $\log r/r$ term for the $Q$-branch.

To mimic the situation valid for electromagnetism and gravity, one can write the charge for the $P$-branch as a flux at null infinity $$\lim_{r \rightarrow \infty} \oint_{S_r} \lambda (x^A) \frac{\phi}{r}( r^2\sqrt{\xbar \gamma} )d^2x  \qquad (\phi^{(0)}_Q = 0) \, ,$$ where $\lambda$ is the function of the angles given by $-\sum_{l,m}\Xi_{lm}^{Q(0)}Y_{lm}(x^{A})$.   One cannot write the charge for the $Q$-branch as a similar flux $\lim_{r \rightarrow} \oint_{S_r} \lambda' (x^A) \frac{\phi}{r} (r^2 \sqrt{\xbar \gamma}) d^2x$ at null infinity because of the $\log r$  divergence.  As we have shown, the fact that this expression is ill-defined for the $Q$-branch does not mean that the $Q$-branch is pathological from the point of view of the charges, but rather, that this naive flux expression is incorrect. A direct and independent derivation of the charges at null infinity by symplectic techniques, taking into account that the induced symplectic form at null infinity involves also $\log r$, is not available here because the charges are not true generators.  This is the reason that we defined them at null infinity from their Hamiltonian expressions at spatial infinity, which take the same uniform form for the two branches, and followed them to null infinity, where the expressions now strikingly differ. 

\vspace{.4cm}

\section{Matching conditions}
\label{Sub:Matching}
To derive the matching conditions of the leading orders of the scalar field between the past of future null infinity and the future of past null infinity, we expand the general solution of the Klein-Gordon equation obtained above in advanced Bondi coordinates $(v, r, x^A)$ and examine its behavior as $v \rightarrow + \infty$.  The procedure is identical with that followed for future null infinity and we directly go to the results.
 
\subsection{Standard matching conditions}
If one imposes the standard parity conditions on the leading order of the fields, one finds that the solution near past null infinity reads
\be
\phi = \frac{\Phi'(v, x^A)}{r} + \mathcal O \left(\frac{\log r}{r^2} \right)\,,
\ee
with
\be
\lim_{v \rightarrow  \infty}  \Phi'(v, x^A) =  \Phi'(x^A)=  \sum_{l,m} \Theta^{P(0)}_{lm} P_{l}^{(\frac12)}(-1) Y_{lm}(x^A)  = \Phi (-x^A)\,. \label{Eq:SchemePv}
\ee
We have thus derived
\be
\lim_{v \rightarrow  \infty}  \Phi'(v, x^A) = \lim_{u \rightarrow - \infty}  \Phi(u, -x^A)\,, \label{Eq:ScalarP}
\ee
where $\Phi$ and $\Phi'$ are the coefficients of the leading $\frac{1}{r}$ order in the expansion of the scalar field near future null infinity and past null infinity, respectively.

The matching given by Eq (\ref{Eq:ScalarP})  is the scalar analog of the matching conditions assumed in \cite{Strominger:2017zoo}.  We have shown that these matching conditions are consequences of our conditions on the initial data: parity conditions on the initial conditions on $t=0$ under the antipodal map $x^A \rightarrow - x^A$  imply matching conditions relating the future of past null infinity and the past of future null infinity ($s \rightarrow -s$, $x^A \rightarrow - x^A$).  Note that the matching  (\ref{Eq:ScalarP}) holds only in the limit (and at leading $\frac{1}{r}$-order)  since $\Phi'(v, x^A) $ is not the antipodal mirror image of $\Phi(u, x^A)$ due to the presence of both $P$ and $Q$ terms in (\ref{Eq:PBranchLeading}).

That the standard matching conditions are consequences of the standard parity-restricted initial data has been derived first in
\cite{Henneaux:2018gfi,Henneaux:2018hdj} (see also \cite{Troessaert:2017jcm,Henneaux:2019yax}).

We can phrase the matching conditions from the point of view of null infinity as follows.   Assume that one directly solves the Klein-Gordon equation near future null infinity by adopting an expansion of the scalar field of the form 
 $$\frac{\Phi(u, x^A)}{r} + \mathcal O \left(\frac{\log r}{r^2} \right)\,, $$
without trying to connect the solution to initial data on Cauchy hypersurfaces.  Assume then that one achieves the same task independently near past null infinity with an expansion 
$$\frac{\Phi'(v, x^A)}{r} + \mathcal O \left(\frac{\log r}{r^2} \right) \, .$$
How should the solutions near future, respectively past, infinity be related if they are to represent the same solution? The answer is that they must be matched as in (\ref{Eq:ScalarP}) for the leading order (the charge).  The initial data corresponding to such a behaviour are furthermore such that $\xbar \phi$ is even and $\xbar \pi$ is odd.

\subsection{Logarithmic matching conditions}
One can equally impose the opposite, non standard parity conditions on the leading orders of the initial data.  In that case, the solution near past null infinity reads
\be
\phi =\frac12  \frac{\log r}{r} \Psi' (v, x^A) + \mathcal O \left(\frac{1}{r} \right) \, ,
\ee
where
\be
\Psi' (v, x^A) =  \Psi'(x^A) =  -\sum_{l,m} \Theta^{Q(0)}_{lm} P_{l}^{(\frac12)}(-1) Y_{lm}(x^A) = - \Psi(-x^A)\,,
\label{Eq:SchemeQv}
\ee
since $Q_0^{(\frac12)}(s) = Q_0^{(\frac12)}(-1+v/r) = - \frac12\log r  + \mathcal O (1)$.   
For the unusual parity conditions,   we have thus established
\be
\lim_{v \rightarrow  \infty}  \Psi'(v, x^A) = - \lim_{u \rightarrow - \infty}  \Psi(u, -x^A)\,, \label{Eq:ScalarQ}
\ee
where $\Psi$ and $\Psi'$ are the coefficients of the leading $\frac{\log r}{r}$ order in the expansion of the scalar field near future null infinity and past null infinity, respectively. 
The matching of the leading order of the scalar field critically involves now a  minus sign\footnote{Because the leading $\mathcal O(\log r/r)$ terms do not depend on $v$ or $u$, the matching actually holds for all values of the advanced and retarded times.}.
 Opposite parity properties at spatial infinity lead to opposite matching conditions between past null infinity and future null infinity for the leading orders of the field (now at $\mathcal O(\log r/r)$ rather than $\mathcal O(1/r)$).

We can again phrase the matching conditions from the point of view of null infinity.   Assume this time that one directly solves the Klein-Gordon equation near future null infinity by adopting an expansion of the scalar field of the form 
 $$\frac12  \frac{\log r}{r} \Psi (u, x^A) + \mathcal O \left(\frac{1}{r} \right)\,, $$
and independently near past null infinity with an expansion 
$$\frac12  \frac{\log r}{r} \Psi' (v, x^A) + \mathcal O \left(\frac{1}{r} \right) \, .$$
How should the solutions near future, respectively past, infinity be related if they are to represent the same solution? The answer is that they must now be matched as in (\ref{Eq:ScalarQ}) for the leading order (the Newman-Penrose charge).  The initial data are furthermore such that $\xbar \phi$ is odd and $\xbar \pi$ is even.

Of course, since we have derived the general solution near future infinity and past infinty to all orders, one can establish from it the matching conditions for the subsequent terms which take, however, a more complicated form because these subleading terms do not have a definite parity.  For relating the fields at the future and past critical sets, the use of the solutions $\widetilde P^{\lambda)}_n$ rather than $\overline P^{\lambda)}_n$ when $n<0$ is more convenient.

\section{Higher dimensions}
\label{SecHighD}

In higher spacetime dimensions $D$, the scalar field and its conjugate momentum decay at spatial infinity as (in Cartesian coordinates)
\be
\phi = \frac{\xbar \phi}{r^{D-3}} + \frac{\phi^{(D-2)}}{r^{D-2}}+ \mathcal O(r^{D-1})\, , \label{Eq:decay1a}%
\ee
and
\be
\pi = \frac{\xbar \pi}{r^{D-2}} + \frac{\pi^{(D-1)}}{r^{D-1}}+ \mathcal O(r^{D})\,. \label{Eq:decay1b}
\ee
This decay is motivated by the behaviour of the elementary solution to the Laplace equation.

As in $D=4$ spacetime dimensions, the general solution of the Klein-Gordon equation can be described in terms of solutions of the Gegenbauer differential equation.  These are again of two types: the $P$-branch, which is even under the hyperboloid antipodal map $s \rightarrow -s$, $x^A \rightarrow - x^A$; and the $Q$-branch, which  is odd\footnote{We note that the authors of \cite{Gasperin:2024bfc} solve the scalar wave equation in the unphysical conformally rescaled Friedrich metric, which is not equivalent for $D>4$ to the scalar wave equation in the physical (Minkowskian) metric since the Ricci scalar of the unphysical metric does not vanish.  This explains the differences. }.

The main new feature with respect to  $D=4$ dimensions, however,  is that parity conditions on the leading orders are no longer necessary to make the action finite in $D>4$ dimensions since the kinetic term is finite in those cases.  Indeed, one has
$$ \int d^{D-1} \pi \dot {\phi} \sim \int dr r^{D-2} r^{-D+3} r^{-D+2} \sim \int dr r^{-D+3} \, ,$$ which does not diverge at infinity whenever $D>4$.  The parity conditions will therefore not be imposed and both the $P$-branch and the $Q$-branch will simultaneously be fully kept. This leads to mixed matching conditions as in electromagnetism \cite{Henneaux:2019yqq}.

In the absence of parity conditions, one gets twice as many conserved charges as in $D=4$: both the even and odd parts of $\xbar \phi$ and of $\xbar \pi$ are conserved and can be kept.  As it will be seen - and expected - the even part of  $\xbar \phi$ and the odd part of  $\xbar \pi$ combine to form a single ``$P$-charge" at null infinity, while the odd part of  $\xbar \phi$ and the even part of  $\xbar \pi$ combine to form the ``$Q$-charge".  These two types of charges exhibit different behaviours at null infinity.

To exhibit these features, we follow exactly the same procedure as in $D=4$ and expand the scalar field in hyperbolic coordinates as 
\begin{equation}
\phi=\sum_{k=0}\eta^{-(k+D-3)}\phi^{(k)}\,,
\end{equation}
to obtain decoupled equations at each order
\begin{eqnarray}
&& -(1-s^{2})^{2}\partial_{s}^{2}\phi^{(k)}+(4-d)(1-s^{2})s\partial_{s}\phi^{(k)} \nonumber \\
&& \qquad \qquad \qquad +(1-s^{2})\xbar D_{A}\xbar D^{A}\phi^{(k)}+(k+d-3)(k-1)\phi^{(k)}=0\,.\qquad
\end{eqnarray}

We then expand each $\phi^{(k)}$ in spherical harmonics as
\begin{equation}
\phi^{(k)}=(1-s^{2})^{\frac{1-k}{2}}\sum_{l,m}\Theta_{lm}^{(k)}Y_{lm}\,,
\end{equation}
with
\begin{equation}
\xbar D_{A}\xbar D^{A}Y_{lm}=-l(l+D-3)Y_{lm}\,,
\end{equation}
to get
\begin{equation}
-(1-s^{2})\partial_{s}^{2}\Theta_{lm}^{(k)}-(2k+D-6)s\partial_{s}\Theta_{lm}^{(k)}+\left(k-l-1\right)\left(k+l+D-4\right)\Theta_{lm}^{(k)}=0\,,\label{eq:a1}
\end{equation}
which we rewrite in commonly adopted notations as
\begin{equation}
(1-s^{2})\partial_{s}^{2}Y_{n}^{(\lambda)}+(2\lambda-3)s\partial_{s}Y_{n}^{(\lambda)}+(n+1)(n+2\lambda-1)Y_{n}^{(\lambda)}=0\,,\label{eq:EqY}
\end{equation}
with 
\begin{equation}
\lambda=k+\frac{D-3}{2}\, ,\qquad  n=l-k\,,
\end{equation}
and $Y_{n}^{(\lambda)} \equiv \Theta_{lm}^{(k)}$.  This is exactly the equation (A.1) of \cite{Henneaux:2018mgn}, which has been thoroughly studied in the appendix A of that reference, to which we refer for technical information.  The parameter $\lambda$ is an integer in odd spacetime dimensions and a half-integer in even spacetime dimensions.  Furthermore $\lambda \geq 1$ when $D \geq 5$.   

The indicial equation for the second order linear differential equation (\ref{eq:EqY}) at the Fuchsian singularities $\pm 1$ is $\alpha^2 - (\lambda - \frac12) \alpha = 0$, with roots $\alpha = 0$ and $\alpha = \lambda - \frac12$. The difference between the two roots  is thus $\lambda - \frac12$ and does not vanish when $D>4$.  It is an integer if and only if $D$ is even.

The structure of the general solution of (\ref{eq:EqY}) parallels that of the $D=4$ case and reads
\begin{equation}
Y_{n}^{(\lambda)}(s)=A\widetilde{P}_{n}^{(\lambda)}(s)+B\widetilde{Q}_{n}^{(\lambda)}(s)\,,
\end{equation}
involving a $P$-branch and a $Q$-branch.
The explicit form of $\widetilde{P}_{n}^{(\lambda)}$ and $\widetilde{Q}_{n}^{(\lambda)}$ depends again on whether $n\geq 0$ or $n<0$ \cite{Henneaux:2018mgn}.
\begin{itemize}
\item For $n \geq 0$, one has
\begin{equation}
\widetilde{P}_{n}^{(\lambda)}=(1-s^{2})^{\lambda-\frac{1}{2}}P_{n}^{(\lambda)}\,,\qquad\widetilde{Q}_{n}^{(\lambda)}=(1-s^{2})^{\lambda-\frac{1}{2}}Q_{n}^{(\lambda)}\,,
\end{equation}
where $P_{n}^{(\lambda)}$ is an ultraspherical polynomial
and $Q_{n}^{(\lambda)}$ is an ultraspherical function of second kind,
leading to the behaviour
\be
\lim_{s \rightarrow 1} \widetilde P_{n}^{(\lambda)} = 0 \, , \qquad \lim_{s \rightarrow 1} \widetilde Q_{n}^{(\lambda)} = \frac{1}{2 \lambda -1} \, \, \quad (\lambda > \frac12)\,,
\ee
near $s=1$ ($ \widetilde Q_{n}^{(\lambda)} $ is associated with the indicial root $\alpha = 0$ while $ \widetilde P_{n}^{(\lambda)} $ is associated with $\alpha =\lambda - \frac12$).  When the spacetime dimension $D$ is even, $\lambda$ is a half-integer and the function  $\widetilde Q_{n}^{(\lambda)}$ involves subleading logarithmic terms but this is not the case when $D$ is odd.  One has
 \be
\widetilde P_n^{(\lambda)}(-s)= (-1)^n \widetilde P_n^{(\lambda)}(s) \, , \qquad \widetilde Q_n^{(\lambda)}(-s)=-(-1)^n \widetilde Q_n^{(\lambda)}(s)\,. \label{Eq:ParityUltra2}
\ee

\item For $n<0$, $\widetilde{Q}_{n}^{(\lambda)}(s)$, which is associated with the root $\alpha = 0$, is a polynomial. When $D$ is even, $\lambda$ is a half-integer and $\widetilde{P}_{n}^{(\lambda)}(s)$ is also a polynomial.  But this will not hold for odd spacetime dimensions where the second indicial root $\alpha =\lambda - \frac12$ is half-integer. We can choose $\widetilde{Q}_{n}^{(\lambda)}(s)$ and $\widetilde{P}_{n}^{(\lambda)}(s)$ so that (\ref{Eq:ParityUltra2}) also holds for negative $n$.  This choice implies that both $\widetilde{Q}_{n}^{(\lambda)}(s)$ and $\widetilde{P}_{n}^{(\lambda)}(s)$ have some finite, non-vanishing  values as $s \rightarrow 1$ \cite{Henneaux:2018mgn}.  Again, one can replace $\widetilde{P}_{n}^{(\lambda)}(s)$ by $\overline{P}_{n}^{(\lambda)}(s)$ that differs from it by a multiple of $\widetilde{Q}_{n}^{(\lambda)}(s)$ so that $\overline{P}_{n}^{(\lambda)}(s)$ vanishes as $(1-s)^{(\lambda - \frac12)}$ for $s \rightarrow 1$.
\end{itemize}

The general solution to the scalar field equation is given
by 
\be
\phi^{(k)}= \phi_P^{(k)}+\phi_Q^{(k)}\,,
\ee
with
\begin{equation}
\phi_P^{(k)}=\sum_{l,m}(1-s^{2})^{\frac{1-k}{2}} \Theta_{lm}^{P(k)}\widetilde{P}_{l-k}^{(k+\frac{d-3}{2})}(s)Y_{lm}\,, 
\end{equation}
and
\begin{equation}
\phi_Q^{(k)}=\sum_{l,m}(1-s^{2})^{\frac{1-k}{2}} \Theta_{lm}^{Q(k)}\widetilde{Q}_{l-k}^{(k+\frac{d-3}{2})}(s)Y_{lm} \, .
\end{equation}
The parity properties are the same as in $D=4$, 
\be
 \phi_P^{(k)}(-s, -x^A) = (-1)^k \phi_P^{(k)}(s, x^A) \,, \qquad \phi_Q^{(k)}(-s, -x^A) = - (-1)^k \phi_Q^{(k)}(s, x^A) \,.
\ee
This yields for $\phi$,
\be
\phi= \phi_P+\phi_Q\,,
\ee
with
\begin{equation}
\phi_P=\sum_{k\geq 0}\sum_{l,m}\eta^{-(k+D-3)}(1-s^{2})^{\frac{1-k}{2}} \Theta_{lm}^{P(k)}\widetilde{P}_{l-k}^{(k+\frac{D-3}{2})}(s)Y_{lm}\,,
\end{equation}
and
\begin{equation}
\phi_Q=\sum_{k\geq0}\sum_{l,m}\eta^{-(k+D-3)}(1-s^{2})^{\frac{1-k}{2}}\Theta_{lm}^{Q(k)}\widetilde{Q}_{l-k}^{(k+\frac{D-3}{2})}(s)Y_{lm}\,.
\end{equation}

Going to Friedrich coordinates, we get
\begin{equation}
\phi_P=(1-s^{2})^{\frac{D-2}{2}}\sum_{k\geq 0}\sum_{l,m}\rho^{-(k+D-3)} \Theta_{lm}^{P(k)}\widetilde{P}_{l-k}^{(k+\frac{D-3}{2})}(s)Y_{lm}\,.
\end{equation}
and
\begin{equation}
\phi_Q=(1-s^{2})^{\frac{D-2}{2}}\sum_{k\geq0}\sum_{l,m}\rho^{-(k+D-3)}\Theta_{lm}^{Q(k)}\widetilde{Q}_{l-k}^{(k+\frac{D-3}{2})}(s)Y_{lm}\,,
\end{equation}
from which we see that $\phi$ goes to zero at least as $(1-s^{2})^{\frac{D-2}{2}}$ at the critical sets where null infinity ``meets" spatial infinity ($s \rightarrow \pm 1$).

For later convenience, we split the $P$-branch as
\begin{eqnarray}
\phi_P&=&\sum_{k\geq 0}\sum_{l\geq k,m}\rho^{-(k+D-3)}(1-s^{2})^{(k+D-3)} \Theta_{lm}^{P(k)}{P}_{l-k}^{(k+\frac{D-3}{2})}(s)Y_{lm}\nonumber \\
&&+ \sum_{k > 0}\sum_{l<k,m}\rho^{-(k+D-3)}(1-s^{2})^{\frac{D-2}{2}} \Theta_{lm}^{P(k)}\widetilde{P}_{l-k}^{(k+\frac{D-3}{2})}(s)Y_{lm}
\end{eqnarray}
In Bondi coordinates, the solution reads
\begin{align}
&\phi_P  =\sum_{k \geq 0}\sum_{l\geq k,m}r^{-(k+D-3)}\Theta_{lm}^{P(k)}P_{l-k}^{(k+\frac{D-3}{2})}(1+\frac{u}{r})Y_{lm} \nonumber \\
 & \qquad+\sum_{k > 0}\sum_{l<k,m}f^{(k,D)}(u,r)\Theta_{lm}^{P(k)}\tilde{P}_{l-k}^{(k+\frac{D-3}{2})}(1+\frac{u}{r})Y_{lm}\,.
\end{align}
and
\be
\phi_Q =\sum_{k\geq 0}\sum_{l,m}f^{(k,D)}(u,r)\Theta_{lm}^{Q(k)}\tilde{Q}_{l-k}^{(k+\frac{D-3}{2})}(1+\frac{u}{r})Y_{lm}\,.
\ee
Here,
\be
f^{(k,D)}(u,r) = r^{-\frac{D-2}{2}}(-2u)^{-\left(k+\frac{D-4}{2}\right)}\left(1+\frac{u}{2r} \right)^{-\left(k+\frac{D-4}{2} \right)}\, .
\ee

Since 
\begin{eqnarray}
&& \hspace{-.4cm}  f^{(k,D)}(u,r) = r^{-\frac{D-2}{2}}(-2u)^{-\left(k+\frac{D-4}{2}\right)} \nonumber \\
&& \qquad \quad +\frac14 \left(k + \frac{D-4}{2}\right) r^{-\frac{D}{2}}(-2u)^{-\left(k+\frac{D-6}{2}\right)} + \mathcal O\left(r^{-\frac{D}{2}-1} \right),
\end{eqnarray}
and
$\tilde{Q}_{l-k}^{(k+\frac{D-3}{2})}(1+\frac{u}{r}) $ is of order one (see Appendix {\bf \ref{App:Gegenbauer}}), 
we see that the leading behaviour of the $Q$-branch at null infinity is $r^{-\frac{D-2}{2}}$ ($D>4$) and involves accordingly fractional powers of $r$ in odd spacetime dimensions.  There are also subdominant logarithms in even spacetime dimensions starting at order $\mathcal O (\log r/r^{D-3})$.

The leading behaviour of the $P$ branch is $r^{-D+3}$ for the terms with $l \geq k$ (in particular, $k=0$), but there are also, for each $k >0$, a finite number of terms behaving as $r^{-\frac{D-2}{2}}$ coming from $l<k$.  

 To understand how the charges exhibited at null infinity match with the  behaviour of the field at null infinity, we consider first the specific cases of $D= 5$ and $D=6$ dimensions.

\subsection{Matching conditions in $D=5$ dimensions}

Using the previous formulas, one finds that the the scalar field behaves near future null infinity as
\begin{equation}
\phi=\frac{1}{r^{3/2}(-2u)^{1/2}}\Psi(u,x^{A})+\frac{1}{r^{2}}\Phi(u,x^{A})+o\left(r^{-2}\right)\,,
\end{equation}
where
\begin{align}
\Psi(u,x^{A}) & =\sum_{k\geq 0}\sum_{l,m}(-2u)^{-k}\Theta_{lm}^{Q(k)}\widetilde{Q}_{l-k}^{(k+1)}(1)Y_{lm} \nonumber \\
& \qquad + \sum_{k> 0}\sum_{l<k,m}(-2u)^{-k}\Theta_{lm}^{P(k)}\widetilde{P}_{l-k}^{(k+1)}(1)Y_{lm}\, , \\
\Phi(u, x^{A}) & =\sum_{l,m}\Theta_{lm}^{P(0)}P_{l}^{(1)}(1)Y_{lm} \equiv \Phi(x^A)\,.
\end{align}
By following a similar procedure, one finds that the solution near past null infinity  is given in advanced null coordinates $(v,r)$ by
\begin{equation}
\phi=\frac{1}{r^{3/2}(2v)^{1/2}}\Psi'(v,x^{A})+\frac{1}{r^{2}}\Phi'(v, x^{A})+o\left(r^{-2}\right)\,,
\end{equation}
where
\begin{align}
\Psi'(v,x^{A}) & =\sum_{k\geq 0}\sum_{l,m}(2v)^{-k}\Theta_{lm}^{Q(k)}\widetilde{Q}_{l-k}^{(k+1)}(-1)Y_{lm} \nonumber \\
& \qquad + \sum_{k> 0}\sum_{l<k,m}(2v)^{-k}\Theta_{lm}^{P(k)}\widetilde{P}_{l-k}^{(k+1)}(-1)Y_{lm}\, , \\
\Phi'(v, x^{A}) & =\sum_{l,m}\Theta_{lm}^{P(0)}P_{l}^{(1)}(-1)Y_{lm} \equiv \Phi'(x^A)\,.
\end{align}

Since only $k=0$ is relevant in the limits $u \rightarrow - \infty$ and $v \rightarrow \infty$, one gets the matching
\be
\lim_{v \rightarrow  \infty}  \Psi'(v, x^A) = - \lim_{u \rightarrow - \infty}  \Psi(u, -x^A)\, , \quad \lim_{v \rightarrow  \infty}  \Phi'(v, x^A) =  \lim_{u \rightarrow - \infty}  \Phi(u, -x^A) \label{Eq:ScalarPQ5}
\ee
The matching is therefore of mixed type and involves a minus sign for the $Q$-branch and a plus sign for the $P$-branch.  Both branches are clearly separated at the critical sets in $D=5$ spacetime dimensions, so that the matching is transparent.  This is similar to what is described in \cite{Henneaux:2019yqq} for the electromagnetic field.

Note the interesting phenomenon that the coefficient of the leading $r^{-\frac32}$ term (i.e., $\Psi$) is now not conserved along null infinity due to the radiation, while the coefficient of the $r^{-2}$ term (i.e., $\Phi$) is strictly conserved.  This is the reverse of what was found in $4$ spacetime dimensions.

\subsection{Matching conditions in $D=6$ dimensions}

The situation is more complicated in $D=6$ spacetime dimensions because there is now an overlap of the subleading $Q$-terms with the leading $P$-terms.  This overlap,  which was excluded in $D=4$ because of the necessary parity conditions in that spacetime dimension, is now permited since parity conditions are not required any more.

Specializing the previous formulas to $D=6$, we get
\begin{align}
\phi_{P} & =\frac{1}{r^{2}}\sum_{k>0}\sum_{l<k,m}(-2u)^{-(k+1)}\Theta_{lm}^{P(k)}\tilde{P}_{l-k}^{(k+\frac{3}{2})}(1)Y_{lm} \nonumber \\
 & \quad+\frac{1}{r^{3}}\sum_{l,m}\Theta_{lm}^{P(0)}P_{l}^{(\frac{3}{2})}(1)Y_{lm} \nonumber \\
 & \quad+\frac{1}{r^{3}}\sum_{k>0}\sum_{l<k,m}(-2u)^{-k}\Theta_{lm}^{P(k)}\left[\frac{1}{4}(k+1)\tilde{P}_{l-k}^{(k+\frac{3}{2})}(1)+p_{l,k}\right]Y_{lm} + \mathcal{O}(r^{-4})\,,
\end{align}
where the constants $p_{l,k}$ are defined through the expansion
\be
\tilde{P}_{l-k}^{(k+\frac{3}{2})}(1+\frac{u}{r})  =\tilde{P}_{l-k}^{(k+\frac{3}{2})}(1)+p_{l,k}\left(-\frac{2u}{r}\right)+\mathcal{O}(r^{-2})\,.
\ee
Their explicit form can be computed recursively from the definition of the Gegenbauer polynomials but will not be needed here since they disappear from the matching conditions of the leading terms.

Similarly, 
\begin{align}
\phi_{Q} & =\frac{1}{r^{2}}\sum_{k\geq0}\sum_{l,m}(-2u)^{-(k+1)}\Theta_{lm}^{Q(k)}\tilde{Q}_{l-k}^{(k+\frac{3}{2})}(1)Y_{lm} \nonumber \\
 & \quad+\frac{\log r}{r^{3}}\sum_{l,m}\frac{1}{8}\Theta_{lm}^{Q(0)}(l+1)(l+2)Y_{lm}  \nonumber \\
 & \quad+\frac{1}{r^{3}}\sum_{l,m}\frac{1}{4}\Theta_{lm}^{Q(0)}\tilde{Q}_{l}^{(\frac{3}{2})}(1)Y_{lm}+\frac{1}{r^{3}}\sum_{l,m}\Theta_{lm}^{Q(0)}g_{l}(u)Y_{lm} \nonumber \\
 & \quad+\frac{1}{r^{3}}\sum_{k>0}\sum_{l,m}(-2u)^{-k}\Theta_{lm}^{Q(k)}\left[\frac{1}{4}(k+1)\tilde{Q}_{l-k}^{(k+\frac{3}{2})}(1)+q_{l,k}\right]Y_{lm}+o(r^{-3})\,,
\end{align}
where 
$g_{l}(u)$ is 
\begin{equation}
g_{l}(u)=-\frac{(l+1)(l+2)}{32}\left(4\log(-u)+4+l(l+3)-4\log2\right)-\frac{1}{2}\partial_{s}R_{l}^{(\frac{3}{2})}(1)\, ,
\end{equation}
and the coefficients $q_{l,k}$, which again drop from the matching conditions, are defined through the expansion
\be
\tilde{Q}_{l-k}^{(k+\frac{3}{2})}(1+\frac{u}{r})  =\tilde{Q}_{l-k}^{(k+\frac{3}{2})}(1)+q_{l,k}\left(-\frac{2u}{r}\right)+o(r^{-1})\,,
\ee
of the functions $\tilde{Q}_{l-k}^{(k+\frac{3}{2})}$. Explicit integral expressions for  $\tilde{Q}_{l-k}^{(k+\frac{3}{2})}$ are given in the Appendix {\bf \ref{App:Gegenbauer}}, from which the  $q_{l,k}$ can be systematically computed. 

Collecting all these results, we find that the scalar field behaves near future null infinity as
\begin{equation}
\phi=\frac{1}{r^2 (-2u)}\Psi(u,x^{A})+\frac{\log r}{r^3} \Xi(u,x^A)+ \frac{1}{r^{3}}\Phi(u,x^{A})+o\left(r^{-3}\right)\,,
\end{equation}
where
\begin{align}
\Psi(u,x^{A}) & =\sum_{k> 0}\sum_{l<k,m}(-2u)^{-k}\Theta_{lm}^{P(k)}\widetilde{P}_{l-k}^{(k+\frac32)}(1)Y_{lm}  \nonumber \\
& \qquad + \sum_{k\geq 0}\sum_{l,m}(-2u)^{-k}\Theta_{lm}^{Q(k)}\tilde{Q}_{l-k}^{(k+\frac{3}{2})}(1)Y_{lm}\, , \\
\Xi(u,x^{A}) & =\sum_{l,m}\frac{1}{8}\Theta_{lm}^{Q(0)}(l+1)(l+2)Y_{lm}\,,\\
\Phi(u, x^{A}) & =\sum_{l,m}\left[\Theta_{lm}^{P(0)}P_{l}^{(\frac{3}{2})}(1)+\frac{1}{4}\Theta_{lm}^{Q(0)}\tilde{Q}_{l}^{(\frac{3}{2})}(1)\right]Y_{lm} +\sum_{l,m}\Theta_{lm}^{Q(0)}g_{l}(u)Y_{lm} \nonumber \\
& \quad+\sum_{k>0}\sum_{l<k,m}(-2u)^{-k}\Theta_{lm}^{P(k)}\left[\frac{1}{4}(k+1)\tilde{P}_{l-k}^{(k+\frac{3}{2})}(1)+p_{l,k}\right]Y_{lm} \nonumber \\
 & \quad+\sum_{k>0}\sum_{l,m}(-2u)^{-k}\Theta_{lm}^{Q(k)}\left[\frac{1}{4}(k+1)\tilde{Q}_{l-k}^{(k+\frac{3}{2})}(1)+q_{l,k}\right]Y_{lm}\,.
\end{align}
Similarly, the solution near past null infinity reads
\begin{equation}
\phi=\frac{1}{r^{2}(2v)}\Psi'(v,x^{A})+\frac{\log r}{r^{3}}\Xi'(v,x^{A})+\frac{1}{r^{3}}\Phi'(v,x^{A})+o\left(r^{-3}\right)\,,
\end{equation}
with
\begin{align}
\Psi'(v,x^{A}) & =\sum_{k>0}\sum_{l<k,m}(2v)^{-k}\Theta_{lm}^{P(k)}\tilde{P}_{l-k}^{(k+\frac{3}{2})}(-1)Y_{lm}+\sum_{k\geq0}\sum_{l,m}(2v)^{-k}\Theta_{lm}^{Q(k)}\tilde{Q}_{l-k}^{(k+\frac{3}{2})}(-1)Y_{lm}\,,\\
\Xi'(v,x^{A}) & =\sum_{l,m}\frac{1}{8}\Theta_{lm}^{Q(0)}(-1)^{l}(l+1)(l+2)Y_{lm}\,,\\
\Phi'(v,x^{A}) & =\sum_{l,m}\left[\Theta_{lm}^{P(0)}P_{l}^{(\frac{3}{2})}(-1)+\frac{1}{4}\Theta_{lm}^{Q(0)}\tilde{Q}_{l}^{(\frac{3}{2})}(-1)\right]Y_{lm}+\sum_{l,m}\Theta_{lm}^{Q(0)}\tilde{g}_{l}(v)Y_{lm} \nonumber \\
 & \quad+\sum_{k>0}\sum_{l<k,m}(2v)^{-k}\Theta_{lm}^{P(k)}\left[\frac{1}{4}(k+1)\tilde{P}_{l-k}^{(k+\frac{3}{2})}(-1)+\tilde{p}_{l,k}\right]Y_{lm} \nonumber \\
 & \quad+\sum_{k>0}\sum_{l,m}(2v)^{-k}\Theta_{lm}^{Q(k)}\left[\frac{1}{4}(k+1)\tilde{Q}_{l-k}^{(k+\frac{3}{2})}(-1)+\tilde{q}_{l,k}\right]Y_{lm}\,,
\end{align}
where
\begin{equation}
\tilde{g}_{l}(v)=\frac{(-1)^{l}(l+1)(l+2)}{32}\left(4\log v+4+l(l+3)-4\log2\right)+\frac{1}{2}\partial_{s}R_{l}^{(\frac{3}{2})}(-1)\,.
\end{equation}

Since only $k=0$ is relevant in the limits $u \rightarrow - \infty$ and $v \rightarrow \infty$, one gets the matching
\be
\lim_{v \rightarrow  \infty}  \Psi'(v, x^A) = - \lim_{u \rightarrow - \infty}  \Psi(u, -x^A)\, , \label{Eq:ScalarPQ6a}
\ee
and
\be
\lim_{v \rightarrow  \infty}  \left(\Phi'(v, x^A) - \Gamma'^Q(v, x^A) \right)=  \lim_{u \rightarrow - \infty}  \left(\Phi(u, -x^A) -\Gamma^Q(u, -x^A) \right) \label{Eq:ScalarPQ6b}
\ee
where $\Gamma^Q(u, x^A)$ is the $Q$-branch contribution to order $r^{-3}$, i.e.,
\begin{align}
\Gamma^Q & = \frac14 \sum_{l,m}\Theta_{lm}^{Q(0)}\tilde{Q}_{l}^{(\frac{3}{2})}(1) Y_{lm} +\sum_{l,m}\Theta_{lm}^{Q(0)}g_{l}(u)Y_{lm} \nonumber \\
& \quad+\sum_{k>0}\sum_{l,m}(-2u)^{-k}\Theta_{lm}^{Q(k)}\left[\frac{1}{4}(k+1)\tilde{Q}_{l-k}^{(k+\frac{3}{2})}(1)+q_{l,k}\right]Y_{lm}\,.
\end{align}
We have also the matching
\be  
\lim_{v \rightarrow  \infty}  \Xi'(v, x^A) =  \lim_{u \rightarrow - \infty}  \Xi(u, -x^A)\, , \label{Eq:ScalarPQ6Log}
\ee
of the coefficients of the $\frac{\log r}{r^{3}}$ term. 

The matching is not only of mixed type involving both signs, but there is now the phenomenon that the $Q$-branch interferes with the $P$-branch through its subleading terms, which must be substracted to get the matching (\ref{Eq:ScalarPQ6b}) of the $P$-branch.  The situation gets even more complicated as one increases the spacetime dimension.

\subsection{Analysis of the conserved quantities}

As in four dimensions, the leading coefficients $\xbar \phi$ and $\xbar \pi$ in the expansion of the scalar field and its conjugate momentum at spatial infinity both define angle-dependent conserved quantities\footnote{The subleading terms can also be combined to form conserved quantities, but these are not expected to survive the introduction of interactions.}.   From the expansion of the fields, one sees that the conserved quantities $\xbar \phi$ and $\xbar \pi$  are completely equivalent to the coefficients $\Theta^{P(0)}_{lm}$ and $\Theta^{Q(0)}_{lm}$.  At null infinity, these coefficients can be most easily read off from the expressions for the fields in the limit $u \rightarrow - \infty$, which provide the initial conditions for the $u$-dependence of the charges at Scri$^+$ (much in the same way as the ``initial value" of the Bondi mass is given at  $u \rightarrow - \infty$, where it equals the ADM mass \cite{Ashtekar:1979xeo,Horowitz:1981uw}).  

Since the expressions at null infinity get more and more intrincate as one goes up in dimension, it is again interesting to follow the evolution of the charges from spatial infinity, where they take a simple and manifestly finite form, to the critical sets.  This is done as in four spacetime dimensions, by going to hyperbolic coordinates and requesting that the charges be $s$-independent.

To that end, we introduce an angle dependent parameter $\epsilon^{(0)}$ and consider the charge
\be
Q[\epsilon^{(0)}]\biggr\rvert_{s=0} = \oint \left(\epsilon^{(0)}\partial_{s}\phi^{(0)}-\partial_{s}\epsilon^{(0)}\phi^{(0)}\right)\sqrt{\xbar\gamma}d^{D-2}x\biggr\rvert_{s=0}\,. \label{Eq:QHighDs0}
\ee
This charge, with no bulk term,  does not have a well-defined canonical action but, as in four dimensions, is useful to keep track of the conserved quantities. We stress again that on $s=0$, $\epsilon^{(0)}$ and $\partial_s\epsilon^{(0)}$ are independent functions of the angles.

The correct expression for the charge $Q(s)$ in a hyperbolic slicing, which reduces to (\ref{Eq:QHighDs0}) at $s=0$ and is conserved, is
given by
\begin{equation}
Q=\oint\left(\epsilon^{(0)}\partial_{s}\phi^{(0)}-\partial_{s}\epsilon^{(0)}\phi^{(0)}+\frac{(D-4)s}{1-s^{2}}\epsilon^{(0)}\phi^{(0)}\right)\sqrt{\xbar\gamma}d^{D-2}x\,.
\end{equation}
The new feature is the additional term containing undifferentiated parameter and field.  It vanishes at $s=0$. One verifies that $\frac{dQ}{ds}=0$ by using  the equation of motion for the scalar field
\begin{equation}
(1-s^{2})^{2}\partial_{s}^{2}\phi^{(0)}+(D-4)(1-s^{2})s\partial_{s}\phi^{(0)}-(1-s^{2})\xbar D_{A}\xbar D^{A}\phi^{(0)}-(D-3)\phi^{(0)}=0\,,
\end{equation}
and imposing that the parameter $\epsilon^{(0)}$ should obey the equation
\begin{equation}
(1-s^{2})^{2}\partial_{s}^{2}\epsilon^{(0)}-(D-4)(1-s^{2})s\partial_{s}\epsilon^{(0)}-(1-s^{2})\xbar D_{A}\xbar D^{A}\epsilon^{(0)}-\left[3-D+(D-4)(1+s^{2})\right]\epsilon^{(0)}=0\,.
\end{equation}

If we make use of the ansatz
\begin{equation}
\epsilon^{(0)}=(1-s^{2})^{\frac{5-D}{2}}\sum_{lm}\Lambda_{lm}Y_{lm}\,,
\end{equation}
the equation becomes
\begin{equation}
(1-s^{2})\partial_{s}^{2}\Lambda_{lm}+(D-6)s\partial_{s}\Lambda_{lm}+(l+D-4)(l+1)\Lambda_{lm}=0\,,
\end{equation}
which is again of the ultraspherical type. The solution then reads
\begin{equation}
\Lambda_{lm}(s)=\Lambda_{lm}^{P(0)}\tilde{P}_{l}^{(\frac{D-3}{2})}(s)+\Lambda_{lm}^{Q(0)}\tilde{Q}_{l}^{(\frac{D-3}{2})}(s)\,.
\end{equation}

We can then decompose the parameter into a $P$-branch and a $Q$-branch as
\begin{equation}
\epsilon^{(0)}=\epsilon_{P}^{(0)}+\epsilon_{Q}^{(0)}\,,
\end{equation}
with
\begin{align}
\epsilon_{P}^{(0)} & =(1-s^{2})^{\frac{5-D}{2}}\sum_{lm}\Lambda_{lm}^{P(0)}\tilde{P}_{l}^{(\frac{D-3}{2})}(s)Y_{lm}\,,\\
\epsilon_{Q}^{(0)} & =(1-s^{2})^{\frac{5-D}{2}}\sum_{lm}\Lambda_{lm}^{Q(0)}\tilde{Q}_{l}^{(\frac{D-3}{2})}(s)Y_{lm}\,.
\end{align}
The charge is then given by
\begin{equation}
Q=Q_{PP}+Q_{PQ}+Q_{QP}+Q_{QQ}\,,
\end{equation}
where
\begin{align}
Q_{PP} & =\oint\left(\epsilon_{P}^{(0)}\partial_{s}\phi_{P}^{(0)}-\partial_{s}\epsilon_{P}^{(0)}\phi_{P}^{(0)}+\frac{(D-4)s}{1-s^{2}}\epsilon_{P}^{(0)}\phi_{P}^{(0)}\right)\sqrt{\xbar\gamma}d^{D-2}x\,,\\
Q_{PQ} & =\oint\left(\epsilon_{P}^{(0)}\partial_{s}\phi_{Q}^{(0)}-\partial_{s}\epsilon_{P}^{(0)}\phi_{Q}^{(0)}+\frac{(D-4)s}{1-s^{2}}\epsilon_{P}^{(0)}\phi_{Q}^{(0)}\right)\sqrt{\xbar\gamma}d^{D-2}x\,,\\
Q_{QP} & =\oint\left(\epsilon_{Q}^{(0)}\partial_{s}\phi_{P}^{(0)}-\partial_{s}\epsilon_{Q}^{(0)}\phi_{P}^{(0)}+\frac{(D-4)s}{1-s^{2}}\epsilon_{Q}^{(0)}\phi_{P}^{(0)}\right)\sqrt{\xbar\gamma}d^{D-2}x\,,\\
Q_{QQ} & =\oint\left(\epsilon_{Q}^{(0)}\partial_{s}\phi_{Q}^{(0)}-\partial_{s}\epsilon_{Q}^{(0)}\phi_{Q}^{(0)}+\frac{(D-4)s}{1-s^{2}}\epsilon_{Q}^{(0)}\phi_{Q}^{(0)}\right)\sqrt{\xbar\gamma}d^{D-2}x\,.
\end{align}

By direct replacement of the solutions we can not only prove that
\begin{equation}
Q_{PP}=Q_{QQ}=0\, ,
\end{equation}
but also find that the non-vanishing charges at the limit $s\rightarrow1$ are given
by
\begin{align}
\lim_{s\rightarrow1}Q_{PQ} & =\oint\left(\sum_{lm}\Lambda_{lm}^{P(0)}P_{l}^{(\frac{D-3}{2})}(1)Y_{lm}\right)\left(\sum_{lm}\Theta_{lm}^{Q(0)}Y_{lm}\right)\sqrt{\xbar\gamma}d^{D-2}x\,,\\
\lim_{s\rightarrow1}Q_{QP} & =-\oint\left(\sum_{lm}\Lambda_{lm}^{Q(0)}Y_{lm}\right)\left(\sum_{lm}\Theta_{lm}^{P(0)}P_{l}^{(\frac{D-3}{2})}(1)Y_{lm}\right)\sqrt{\xbar\gamma}d^{D-2}x\,.
\end{align}
As in four dimensions, the $P$-branch (corresponding to the even part of $\xbar \phi$ and the odd part of $\xbar \pi$ under the sphere antipodal map) and the $Q$-branch   (corresponding to the odd  part of $\xbar \phi$ and the even part of $\xbar \pi$) are naturally separated.

\section{Conclusions}
\label{sec:Conclusions}

In this paper, we have investigated the behaviour  of a massless scalar field with long range initial data near the past of future null infinity and the future of past null infinity (``critical surfaces").  The adopted decay at spatial infinity was the one of the elementary solution of the Laplace equation.  The behaviour found at null infinity is very different from the one that characterizes initial data with compact support and involves non-analytic functions of $1/r$, such as logarithms or fractional powers.  This is because the critical surfaces correspond then to Fuchsian singularities of the differential equations controlling the asymptotic behaviour, while spatial infinity defines a regular point.

Two of the central issues that we have considered are the matching conditions and the expression of the conserved charges at the critical surfaces. 

A key feature in the derivation of the matching conditions is the symmetry of the asymptotic differential equations written in hyperbolic coordinates under the reversal of the hyperbolic time $s \rightarrow - s$, which exchanges future and past.  As a result, the solutions, the angle dependence of which is decomposed in  spherical harmonics, split into a $P$-branch and a $Q$-branch.  The $P$-branch is even under the hyperbolic antipodal map (i.e., the time reversal $s \rightarrow - s$ combined with the sphere antipodal map) and the $Q$-branch is odd.  On the initial hypersurface $s=0$, the hyperbolic antipodal map reduces to the sphere antipodal map.  This explains why parity conditions on the initial conditions have a direct impact on the form of the matching conditions relating $s=1$ and $s = -1$\footnote{The connection between parity conditions on the leading order of the initial data and matching conditions has been rederived by different route in \cite{Mohamed:2023jwv}, confirming the earlier results of \cite{Henneaux:2018hdj}.}. 

The non-analytic terms at null infinity originate from the $Q$-branch.  We have explicitly shown that the conserved quantities remain well-defined and finite even when the $Q$-branch is turned on.  These charges  are however not given by expressions at null infinity that are the same as the ones for the charges related to the $P$-branch.  Only the latter can be written as familiar fluxes at null infinity.  The reason that the $Q$-branch non-analytic terms do not raise difficulties for the charges at null infinity is that  the $Q$-branch is  perfectly all right at spatial infinity (analytic in $1/r$), where it merely differs from the $P$-branch by somewhat ``insignificant" (at spatial infinity) parity properties.   

The consideration of long range initial data for the scalar field is motivated by  electromagnetism and gravity, to which we shall turn in future papers.  We shall discuss the matching conditions not only for the gauge invariant field strengths but also for quantities that are strictly invariant under proper gauge transformations but not invariant under improper ones, like the Goldstone bosons of the asymptotic symmetries. We shall also study the conserved charges with due consideration of the $Q$-branch, which complicates the structure at null infinity. The charges are more interesting in the electromagnetic and gravitational cases because they are then true canonical generators: they actually generate the improper gauge symmetries. Symplectic techniques can thus be used in their analysis. We shall also treat the higher-dimensional cases, where Hamiltonian methods have already been developed \cite{Henneaux:2019yqq,Fuentealba:2023huv,Fuentealba:2021yvo,Fuentealba:2022yqt,Fuentealba:2023fwe}. 

To a large extent, many of the ideas and techniques of this paper have been derived earlier in the references \cite{Henneaux:2018cst,Henneaux:2018gfi,Henneaux:2018hdj,Henneaux:2018mgn,Henneaux:2019yqq,Henneaux:2019yax} of which our work is the continuation.  One of the novel developments reported here is the explicit treatment of the $Q$-branch in the matching conditions, which are more subtle since one has to deal with the leading logarithmic term in four spacetime dimensions.  We have also clarified why the non-analytical terms characteristic of the $Q$-branch are harmless for the charges, contrary to incorrect views that one might perhaps hold.  

Also for future work is the derivation of the matching conditions when logarithmic supertranslations or angle-dependent logarithmic $u(1)$ gauge transformations are included \cite{Fuentealba:2022xsz,Fuentealba:2023syb,Fuentealba:2023rvf,Fuentealba:2023huv}.  The physical implications of the $Q$-branch, in particular for the soft theorems, should be explored too.

\section*{Acknowledgments}

It is a pleasure to thank C\'edric Troessaert for discussions in the early stages of this work. O. F. is grateful to the Coll\`ege de France and the Universit\'e Libre de Bruxelles for kind hospitality while this article was completed.  This work was partially supported by  FNRS-Belgium (convention IISN 4.4503.15), as well as by funds from the Solvay Family. O.F. holds a Marina Solvay fellowship.

\appendix

\section*{Appendices}

\section{Coordinate systems
\label{app-decomp}}
Our conventions and notations are as follows.
The Minkowski metric in $D$ spacetime dimensions reads
\begin{equation}
d \sigma^2 = \eta_{\mu \nu} dx^\mu dx^\nu =- dt^2 + \sum_{i=1}^{3} \left(dx^i\right)^2
\end{equation}
($x^0 = t$, $c=1$) or in polar coordinates,
\begin{equation}
d \sigma^2 = - dt^2 + dr^2 + r^2 d\omega_{D-2}^2 \, ,
\end{equation}
where
\begin{equation}
d \omega_{D-2}^2 = \xbar \gamma_{AB} dx^A dx^B \, ,
\end{equation}
gives the line element on the unit round $(D-2)$-sphere.  The metric coefficients $\xbar \gamma_{AB}$ depends only on the sphere coordinates $x^A$ ($A = 1, \cdots, D-2$).

\subsection{Hyperbolic coordinates}
The hyperbolic coordinates are defined in the region $r > \vert t \vert$ by
\be 
\eta = \sqrt{-t^2 + r^2}, \; \; \; s = \frac{t}{r} 
\ee
(angles unchanged). The Minkowski metric becomes
\be 
d \eta^2 + \eta^2 h_{ab} dx^a dx^b, \; \; \; \; (x^a) \equiv (s, x^A)
\ee
with 
\be
h_{ab} dx^a dx^b = - \frac{1}{\left(1-s^2\right)^2} ds^2 + \frac{\xbar \gamma_{AB}}{1-s^2} dx^A dx^B \, .
\ee
The inverse coordinate transformation is given by
\begin{equation}
    t = \eta \frac {s}{\sqrt{1-s^2}}, \quad r = \eta \frac
    {1}{\sqrt{1-s^2}}. \end{equation}

\subsection{Friedrich coordinates}
The Friedrich are obtained by replacing $\eta$ by $\rho$ defined as
\be
\rho = \eta \left(1 - s^2 \right)^{\frac12}, \quad 0<\rho<\infty, \quad -1 <s <1  \, .
\ee
In these coordinates, the Minkowskian line element becomes
\be
d \Sigma^2 = \frac{\rho^2}{\left(1 - s^2\right)^2} \left(\frac{(1-s^2)}{\rho^2} d \rho^2 + 2 s\, \rho^{-1}\,  ds\,  d\rho -  ds^2 +  d \Omega^2 \right) 
\ee
and is conformal to the metric
\be
d \tilde \Sigma ^2 = \frac{(1-s^2)}{\rho^2} d \rho^2 + 2 s\, \rho^{-1}\,  ds\,  d\rho -  ds^2 +  d \Omega^2 \, .
\ee
Spatial infinity is blown up as in hyperbolic coordinates and characterized by $\rho = \infty$, $s \in (-1, 1)$, with different boosted hyperplanes cutting it at different values of $s$.  In that representation, spatial infinity is in fact a timelike cylinder.  While still located at an infinite distance away in the rescaled metric $d \tilde \Sigma ^2$, its induced metric is regular and given by
$
 -  ds^2 +  d \Omega^2 \, .
$

The coordinate $s$ still behaves as $s \rightarrow +1$ (respectively, $-1$) as one goes to $\mathscr{I^+}$ (respectively $\mathscr{I^-}$), but the new coordinate $\rho$ assumes now values that encodes the information on ``where'' one reaches $\mathscr{I^+}$ (respectively $\mathscr{I^-}$).  Specifically, for $\mathscr{I^+}$, one finds,
\be
\rho \rightarrow 2 \vert b \vert \in (0, \infty)  \, .
\ee
Thus, future null infinity $\mathscr{I^+}$ is given by $s =1$, $\rho \in (0,\infty)$.  Similarly, past null infinity $\mathscr{I^-}$ is given by $s =-1$, $\rho \in (0,\infty)$. 
The limitation mentioned in Subsection {\bf \ref{SubSec:FB}} has therefore been eliminated. 

The rescaled metric reduces to the degenerate metric
$
  d \Omega^2
$
on $\mathscr{I^+}$ ($s=1$) and $\mathscr{I^-}$ ($s=-1$).  The boundaries $\rho = \infty$ of $\mathscr{I^+}$ ($s=1$) and $\mathscr{I^-}$ ($s=1$) are denoted $\mathcal{I}^+ \equiv \mathscr{I^+_-}$ and $\mathcal{I}^-\equiv \mathscr{I^-_+}$.  They are the critical surfaces (spheres).  They are also the upper and lower boundaries of the cylinder representating spatial infinity.  Matching conditions connect limiting values of fields at the critical surfaces.  

Note that $s = 1$, $\rho \rightarrow \infty$ corresponds to going first to $\mathscr{I^+}$ ($s=1$, $\rho >0$) and then taking the limit to the past of $\mathscr{I^+}$ (i.e., going to spatial infinity ``from above'' along null infinity).  Similarly, $s = -1$, $\rho \rightarrow \infty$ corresponds to going first to $\mathscr{I^-}$ ($s=-1$, $\rho >0$) and then taking the limit to the future of $\mathscr{I^-}$ (i.e., going to spatial infinity ``from below'' along null infinity).  Note also the fact that increasing $\rho$ corresponds to going to spatial infinity in both cases - hence to the past for $\mathscr{I^+}$ and to the future for $\mathscr{I^-}$. 

Conversely, the limit $\rho \rightarrow \infty$, $s \in (-1,1)$ followed by $s \rightarrow 1$ (respectively $-1$) corresponds to going to future null infinity (respectively past null infinity) from spatial infinity.  

\section{Gegenbauer functions of the second kind}
\label{App:Gegenbauer}

As we have seen in the text, the relevant differential equations controlling the asymptotic dynalmics of the scalar field are
\begin{equation}
(1-s^{2})\partial_{s}^{2}Y_{n}^{(\lambda)}+(2\lambda-3)s\partial_{s}Y_{n}^{(\lambda)}+(n+1)(n+2\lambda-1)Y_{n}^{(\lambda)}=0\,,
\end{equation}
with 
\begin{equation}
\lambda=k+\frac{D-3}{2}\, ,\qquad  n=l-k\,,
\end{equation}
(see (\ref{eq:EqY})).  

For $n\geq 0$, the solution can be written as
\begin{equation}
Y_{n}^{(\lambda)}(s)=A\widetilde{P}_{n}^{(\lambda)}(s)+B\widetilde{Q}_{n}^{(\lambda)}(s)\,,
\end{equation}
where
\begin{equation}
\widetilde{P}_{n}^{(\lambda)}=(1-s^{2})^{\lambda-\frac{1}{2}}P_{n}^{(\lambda)}\,,\qquad\widetilde{Q}_{n}^{(\lambda)}=(1-s^{2})^{\lambda-\frac{1}{2}}Q_{n}^{(\lambda)}\,.
\end{equation}
Here, $P_{n}^{(\lambda)}$ is an ultraspherical (Gegenbauer) polynomial
and $Q_{n}^{(\lambda)}$ is an ultraspherical function of second kind.  As recalled in the appendix A of \cite{Henneaux:2018mgn} these functions can be constructed using the recurrence relation
\be
n Q_{n}^{(\lambda)}(s) = 2(n+ \lambda - 1) s Q_{n-1}^{(\lambda)}(s) - (n+2 \lambda - 2) Q_{n-2}^{(\lambda)}(s) \, ,
\ee
with starting point
\be
Q_{0}^{(\lambda)}(s) = \int_0^s \left(1-x^2\right)^{-\lambda - \frac12} dx \, , \qquad Q_{1}^{(\lambda)}(s) = 2 \lambda s Q_{0}^{(\lambda)} (s) - \left(1-s^2\right)^{-\lambda + \frac12} \, .
\ee

To be more explicit on the $s$-dependence of $Q_{0}^{(\lambda)}(s)$, we note that if we set $\alpha=\cos^{-1}x$, the integrals become
\begin{equation}
\int_0^s \left(1-x^{2}\right)^{-p}dx=\int_{\arccos s}^{\frac{\pi}{2}}(\sin \alpha)^{-(2p-1)}d\alpha\,,\label{eq:Integral1}
\end{equation}
with $p = \lambda + \frac12$ integer (even dimensions) or half-integer (odd dimensions).
Let us set 
\begin{equation}
m=2p-1\,, \qquad J_{m}=\int(\sin\alpha)^{-m}\, d\alpha\,.
\end{equation}
For $p$ integer, $m$ is odd, while for $p$ half-integer, $m$ is
even. 

The above integral can be written as
\begin{align}
\int(\sin\alpha)^{-m}d\alpha & =\int(\sin\alpha)^{-(m-1)}\sin\alpha d\alpha \nonumber \\
 & =-(\sin\alpha)^{-(m-1)}\cos \alpha-(m-1)\int(\cos\alpha)^{2}(\sin\alpha)^{-m}d\alpha \nonumber\\
 & =-(\sin\alpha)^{-(m-1)}\cos\alpha-(m-1)\int(\sin\alpha)^{-m}d\alpha+(m-1)\int (\sin\alpha)^{-(m-2)}d\alpha\,.
\nonumber \end{align}
From the latter, we obtain that
\begin{equation}
m\int(\sin\alpha)^{-m}d\alpha=-(\sin\alpha)^{-(m-1)}\cos\alpha+(m-1)\int(\sin\alpha)^{-(m-2)}d\alpha\,.
\end{equation}
or
\begin{equation}
J_{m}=-\frac{1}{m}(\sin\alpha)^{-(m-1)}\cos\alpha+\frac{(m-1)}{m}J_{m-2}\,.\label{eq:Recursion}
\end{equation}

For the integrals with $p$ half-integer in \eqref{eq:Integral1},
we compute the lowest case. Explicit computation shows that
\begin{equation}
J_{2}(s)=\int_{\arccos s}^{\frac{\pi}{2}} (\sin\alpha)^{-2}d\alpha=\frac{s}{\sqrt{1-s^{2}}}\,.
\end{equation}
Then, by virtue of the recursion integral
formula \eqref{eq:Recursion}, $J_{m}$ will not possess logarithmic
terms for all $m\geq4$ even ($p$ half-integer), which corresponds
to the case of  odd dimensions.

For the integrals with $p$ integer in \eqref{eq:Integral1}, we only
must compute the lowest case to show the presence of logarithmic terms
in this class of integrals. Explicit computation shows that
\begin{equation}
J_{3}=\int_{\arccos s}^{\frac{\pi}{2}}(\sin\alpha)^{-3}d\alpha=\frac{x}{2(1-x^{2})}+\frac{1}{4}\log\left(\frac{1+x}{1-x}\right)\,.
\end{equation}
Thus, $J_{m}$ will possess logarithmic terms for all $m\geq5$ odd
($p$ integer), which only happens in even dimensions.  

The appearance of fractional powers of $r$ (and no logarithm) in odd dimension, as well as the appearance of logarithms in even dimensions, of course perfectly agrees with the general theory of linear second order differential equations near Fuchsian singularities and the computation of the roots of the indicial equation.


\end{document}